\documentclass[useAMS,usenatbib]{mn2e} \usepackage{url}
\usepackage{color}
\usepackage{pdfpages}
\usepackage[colorlinks=true,citecolor=blue]{hyperref}
\usepackage{graphicx,graphics,color}
\usepackage{float}
\usepackage{rotating}
\usepackage{amsbsy,amsmath}
\usepackage{amssymb}
\usepackage{natbib}
\usepackage{gensymb}

\title[Classical radio galaxies with uGMRT and MeerKAT]{A new look at old friends. I. Imaging classical radio galaxies with uGMRT and MeerKAT}

\author[Fanaroff, et~al.]{Bernie Fanaroff$^{1}$,
Dharam V. Lal$^{2}$\thanks{E-mail: dharam@ncra.tifr.res.in},
Tiziana Venturi$^{3}$,
Oleg M. Smirnov$^{4,1}$, \newauthor
Marco Bondi$^{3}$,
Kshitij Thorat$^{5,6}$,
Landman H. Bester$^{1,4}$,
Gyula I. G. J\'ozsa$^{1,4,7}$, \newauthor
Dane Kleiner$^{8}$,
Francesca Loi$^{8}$,
Sphesihle Makhathini$^{9}$,
and Sarah V. White$^{4}$
\\
$^{1}$South African Radio Astronomy Observatory, 2 Fir Street, Black River Park, Observatory, Cape Town 7925, South Africa \\
$^{2}$National Centre for Radio Astrophysics - Tata Institute of Fundamental Research, Post Box 3, Ganeshkhind P.O., Pune 411007, India \\
$^{3}$INAF - Istituto di Radioastronomia, via Gobetti 101, I-40129 Bologna, Italy \\
$^{4}$Department of Physics and Electronics, Rhodes University, PO Box 94, Makhanda 6140, South Africa \\
$^{5}$Department of Physics, University of Pretoria, Hatfield, Pretoria, 0028, South Africa \\
$^{6}$Inter-University Institute for Data Intensive Astronomy, Dept. of Astronomy, University of Cape Town, Rondebosch, 7701, South Africa \\
$^{7}$Argelander-Institut f\"ur Astronomie, Auf dem H\"ugel 71, D-53121 Bonn, Germany \\
$^{8}$ INAF - Osservatorio Astronomico di Cagliari, via della Scienza 5, I-09047 Selargius, Italy \\
$^{9}$School of Physics, University of the Witwatersrand, 1 Jan Smuts Avenue, Johannesburg, South Africa \\
}

\date{Accepted XXX. Received YYY; in original form ZZZ}

\pubyear{2018}

\begin{document}
\pagerange{\pageref{firstpage}--\pageref{lastpage}}
\maketitle

\label{firstpage}

\begin{abstract}

We have undertaken a systematic study of FR\,I and FR\,II radio galaxies with the upgraded Giant Metrewave Radio Telescope (uGMRT) and MeerKAT.  The main goal is to explore whether the unprecedented few $\mu$Jy sensitivity reached in the range 550--1712 MHz at the resolution of $\sim4^{\prime\prime}-7^{\prime\prime}$ reveals new features in the radio emission which might need us to revise our current classification scheme for classical radio galaxies.  In this paper we present the results for the first set of four radio galaxies, {\it i.e.} 4C\,12.02, 4C\,12.03, CGCG\,044--046 and CGCG\,021--063.
The sources have been selected from the 4C sample with well-defined criteria, and have been imaged with the uGMRT in the range 550--850 MHz (band 4) and with the MeerKAT in the range 856--1712 MHz (L-band). Full resolution images are presented for all sources in the sample, together with MeerKAT in-band spectral images. Additionally, the uGMRT-MeerKAT spectral image and MeerKAT L-band polarisation structure are provided for CGCG\,044--046.
Our images contain a wealth of morphological details, such as filamentary structure in the emission from the lobes, radio emission beyond the hot-spots in three sources, and misalignments. We briefly discuss the overall properties of CGCG\,044--046 in the light of the local environment as well, and show possible restarted activity in 4C\,12.03 which needs to be confirmed. 
We conclude that at least for the sources presented here, the classical FR\,I/FR\,II morphological classification still holds with the current improved imaging capabilities, but the richness in details also suggests caution in the systematic morphological classification carried out with automatic procedures in surveys with poorer sensitivity and angular resolution.

\end{abstract}


\section{Introduction}
\label{intro}

The morphology of extragalactic radio sources of high and low luminosity, set out by \citet{FanaroffRiley} has suggested the framework for the study of the physics, origin, and evolution of extragalactic radio sources.
After more than 46 years, the classification of extended extragalactic radio sources in FR\,I and FR\,II types is still used to separate low power and high power radio galaxies respectively. However, our knowledge of extragalactic radio sources has improved considerably since then.

\begin{table*}
\begin{center}
\caption{Source properties.}
\begin{tabular}{lccclccl}
\hline
\text{Name} & \text{Type} & \text{R.A.} & \text{Dec.} & \multicolumn{1}{c}{\text{$z$}} & \text{logP$_{\rm 1.4~GHz}$} & \text{kpc/$^{\prime\prime}$} & Alt. name \\
 & & \multicolumn{2}{c}{(J2000)} & & W~Hz$^{-1}$ & & \\
\hline
4C\,12.02    & FR\,II          & 00 04 50.2 & $+$12 48 40 & 0.143    & 26.06 & 2.530 & G4Jy\,7\\
4C\,12.03    & FR\,II          & 00 09 52.6 & $+$12 44 05 & 0.156    & 25.82 & 2.721 & G4Jy\,18\\
CGCG044$-$046 (4C\,07.32) & FR\,I/II (WAT$^{\dagger}$) & 13 16 17.0 & $+$07 02 47 & 0.050145 & 25.08 & 0.986 & G4Jy\,1060 \\
CGCG021$-$063 (4C\,00.56) & FR\,I/II       & 15 16 40.2 & $+$00 15 02 & 0.052489 & 25.25 & 1.030 & G4Jy\,1238 \\
\hline
\end{tabular}
\end{center}
\label{tab:sample}
\begin{flushleft}
${\dagger}$: Wide angle tail (WAT) radio galaxy, see also Sec.~\ref{intro}. \\
The alternative names of radio sources are from the GLEAM 4-Jy sample \citep{2020PASA...37...18W,2020PASA...37...17W}.
\end{flushleft}
\end{table*}

It is known that FR\,I radio galaxies (radio powers typically below $10^{24}$ W~Hz$^{-1}$ at 1.4 GHz) with their symmetric prominent jets and lobes, are associated with red passive galaxies, {\it i.e.} evolved galaxies which usually show little to no evidence of nuclear activity in other bands of the spectrum. More recently \citep[see][for a recent review]{BestHeckman,HardcastleCroston}, these have been associated with low excitation emission-line galaxies \citep[LEGs,][]{HeckmanBest} at moderate luminosities.
Deviations from the straight (180$^{\circ}$) FR\,I morphology is common for optical hosts in galaxy clusters, where radio galaxies may form spectacular tails as the radio jets are exposed to a combination of effects, such as galaxy motion through the intracluster medium (ICM), a.k.a., tailed radio sources \citep{RudnickOwen1976,ODeaOwen1985}, bulk motions of the ICM ({\it i.e.} ``cluster weather'' \citep{Burnsetal}), shocks in the ICM \citep{Noltingetal} and other phenomena related to the formation of clusters \citep[see the review by][for a comprehensive overview]{BrunettiJones2014}.
On the other hand, FR\,II radio galaxies (radio powers above $10^{24}$ W~Hz$^{-1}$ at 1.4 GHz) are associated both with quasars and galaxies and the optical host usually shows other indicators of nuclear activity, such as high excitation emission lines \citep[HEGs,][]{HeckmanBest}; see also \citep[][for a recent review]{HardcastleCroston}.
The jets of these sources are often faint (sometimes barely visible) and asymmetric, and the radio spectral information indicates that the lobes are the result of back-flow from the hot spots.

A lot of work has been done to understand the different behaviour of jets in FR\,Is and FR\,IIs \citep[][and references therein]{Bicknelletal,LaingBridle2012}.
Very-long baseline interferometric studies of samples of FR\,I and FR\,II radio galaxies show that jets are relativistic at their origin in both classes of sources \citep{Venturietal1995,Giovanninietal}; however jets in FR\,I decelerate closer to the core compared to FR\,II, most likely due to differences in some combination of jet power and propagating medium.
It is commonly stated that FR\,I preferentially reside in dense environments, such as clusters and groups of galaxies, and FR\,II are found in less dense environments \citep{Hardcastle2005}, but the most famous FR\,II radio galaxy, Cygnus-A, is at the centre of a galaxy group \citep{BM70}.
FR\,IIs appear to populate the Universe to much larger redshifts. This is partly due to selection effects related to the sensitivity and resolution of the radio interferometers used so far in large surveys. Imaging of large regions of the sky with the current generation of radio interferometers, such as LOFAR (the LOw Frequency ARray), ASKAP (Australia SKA Pathfinder) and MeerKAT clearly shows a wealth of extended radio galaxies with a broad distribution of size, flux density and distance, which might change the paradigm of redshift distribution of FR\,Is and FR\,IIs \citep{Croston2019,Mingoetal, Croston2017}.

\begin{table*}
\caption{Log of the observations using uGMRT and MeerKAT arrays.}
\begin{center}
\begin{tabular}{lcccccccc}
\hline
Source &  Obs. Date & Flux cal. & Phase cal.& Bandwidth & Ch.-width & t$_{\rm int.}$ & FWHM & \textsc{rms} \\
   &      & & &  (MHz) &   (kHz) & (hour) & ($^{\prime\prime}\times^{\prime\prime}, ^{\circ}$)& ($\mu$Jy~b$^{-1}$) \\
\hline\noalign{\smallskip}
\multicolumn{9}{l}{uGMRT array} \\
4C\,12.02     & 2019-09-05 & 3C\,48 & 0054$-$035& 300 & 48.83 &  3.0 &5.17$\times$3.23, ~9.18 & $\sim 30$ \\
4C\,12.03     & 2019-09-17 & 3C\,48 & 0054$-$035& 300 & 48.83 &  2.9 &5.03$\times$3.54, 40.85 & $\sim 21$ \\
CGCG\,044-046 & 2019-05-20 &3C\,147 & 1419$+$064& 200 & 24.41 &  3.4 &4.98$\times$3.76, 53.56 & $\sim 15$ \\
CGCG\,021-063 & 2019-05-26 &3C\,286 & 1419$+$064& 300 & 48.83 &  3.1 &6.99$\times$3.25, 16.38 & $\sim 150$ \\
\hline\noalign{\smallskip}
\multicolumn{9}{l}{MeerKAT array} \\
4C\,12.02     & 2019-05-25 & PKS\,0408$-$65 & J0022$+$0014&  856  & 208.98 & 3.0  &8.65$\times$5.51, 161.95 & $\sim 13$ \\
4C\,12.03     & 2019-05-25 & PKS\,0408$-$65 & J0022$+$0014&  856  & 208.98 & 2.7  &8.52$\times$5.44, 164.56 & $\sim 8$ \\
CGCG\,044-046 & 2019-05-13 & B1934$-$638 & J1347$+$1217&  856  & 208.98 & 3.0 \\
              & 2019-09-15 & B1934$-$638 & J1347$+$1217&  856  & 208.98 & 2.7 &7.51$\times$6.27, 159.50 & $\sim 10$ \\
CGCG\,021-063 & 2019-05-13 & B1934$-$638 & J1512$-$0906&  856  & 208.98 & 3.0  \\
             & 2019-09-15 & B1934$-$638 & J1512$-$0906&  856  & 208.98 &  3.2 &8.31$\times$6.01, 161.77 & $\sim 13$ \\
\hline
\end{tabular}
\end{center}
\label{tab:obs-log}
\end{table*}

A small fraction of FR\,I and FR\,II radio galaxies \citep[of the order of a few percent,][]{1999MNRAS.309..100I,2016MNRAS.461.3165B}, shows Mpc-scale extent, posing the problem of the energy supply to the lobes, and a number of them show signs of restarted activity 
\citep{2020A&A...635A...5D,2019ApJ...875...88B,2019A&A...622A..13M}.
Equally important is a new class of low power radio galaxies that has been recently characterised, the so-called FR\,0 \citep{Baldietal}. The radio power of these sources is typical of FR\,Is. However, they are compact on the scale of a few arcseconds. High sensitivity observations at high angular resolution show that a fraction of them have double-sided jets on very small angular scales, but the majority remain compact \citep{2020A&A...642A.107C}. The nature of these sources and how they fit into the overall classification of radio galaxies is still uncertain.

A wealth of amazing new images has been collected over the past few years with the current generation of radio interferometers, such as LOFAR, JVLA, MeerKAT and uGMRT, whose much improved sensitivity at arcsecond resolution over at least two orders of magnitude in frequency is revealing new features in the radio emission. A few remarkable examples are the amazing tails of NGC\,326 detected with LOFAR \citep{Hardcastlen326}, the low surface brightness emission surrounding several radio galaxies at the centre of groups \citep[for instance NGC\,741,][]{Schnellenbergetal}, the filamentary structure within and outside the lobes of Fornax A \citep{Maccagnietal}, and the filaments of ESO\,137-006 \citep{Ramatsokuetal}.

The imaging capabilities available nowadays thus allow us to throw new light on our understanding of the radio galaxy phenomenon and address questions that have so far remained unanswered. In particular: (a) Why are the morphological classification in terms of FR\,I and FR\,II and the radio power so closely linked, and what do intermediate objects tell us? (b) Which mechanism produces radio loudness in the form of FR\,I or FR\,II in galaxies, what is the duty cycle of radio nuclear activity, and what triggers it?
(c) Does the sharp classification in FR\,I and FR\,II types still hold when we improve our imaging capabilities to the level which can be currently achieved? (d) Do we need more morphological classes?

To address at least the third and fourth of these questions, we started an imaging project of classical radio galaxies with uGMRT in 550--850\,MHz (band-4) and MeerKAT in 856--1712\,MHz (L-band), to explore their radio morphology with unprecedented sensitivity at arcsecond resolution, and study their spectral structure in the frequency range 0.5--1.7 GHz.
Here we present our results of the pilot study for the first set of four radio galaxies. 

The paper is organised as follows. We present our target selection, observations and data analysis in Sec.~\ref{sec.data-red}, followed by the radio morphology in Sec.~\ref{sec.morph-spec}.
In Sec.~\ref{spectral-analysis}, we show the integrated radio spectra and spectral imaging for our sources. 
In Sec.~\ref{sec.sum-conc} we discuss our results, and concluding remarks are given in  Sec.~\ref{conc-rem}.
We assume a $\Lambda$CDM cosmology with $\Omega_{\rm m}$ = 0.27, $\Omega_{\Lambda}$ = 0.73, and H$_0$ = 70 km s$^{-1}$ Mpc$^{-1}$.
We define spectral index, $\alpha$ as, S$_\nu$ $\propto$ $\nu^\alpha$; where S$_\nu$ is the flux density at frequency $\nu$.
Throughout, positions are given in J2000 coordinates.

\section{Target selection and observations}
\label{sec.data-red}

We selected a sample of FR\,I and FR\,II radio galaxies from the 4C catalogue \citep[$-$7$^\circ$ $<$ $\delta$ $<$ 80$^\circ$;][]{PilkingtonScott} which meet the following selection criteria:
\begin{itemize}
\item[(i)] The source is hosted by a detected optical galaxy with spectroscopically determined redshift $z$ in the range between 0.05 and 0.20. This ensures (i) the detection of Mpc scale extended
emission with uGMRT band-4, and (ii) a similar fraction of FR\,Is and FR\,IIs;
\item[(ii)] The target source is in the declination range $-$10$^{\circ}$ and $+$20$^{\circ}$ to ensure proper visibility and comparable ($u,v$)-coverage with both uGMRT and MeerKAT arrays at respective bands; and
\item[(iii)] A clear double radio morphology is imaged at the 45$^{\prime\prime}$ angular resolution of the NRAO VLA Sky Survey \citep[NVSS:][]{Condonetal}.
\end{itemize}

\begin{figure*}
\begin{center}
\begin{tabular}{rl}
\includegraphics[height=6.5cm]{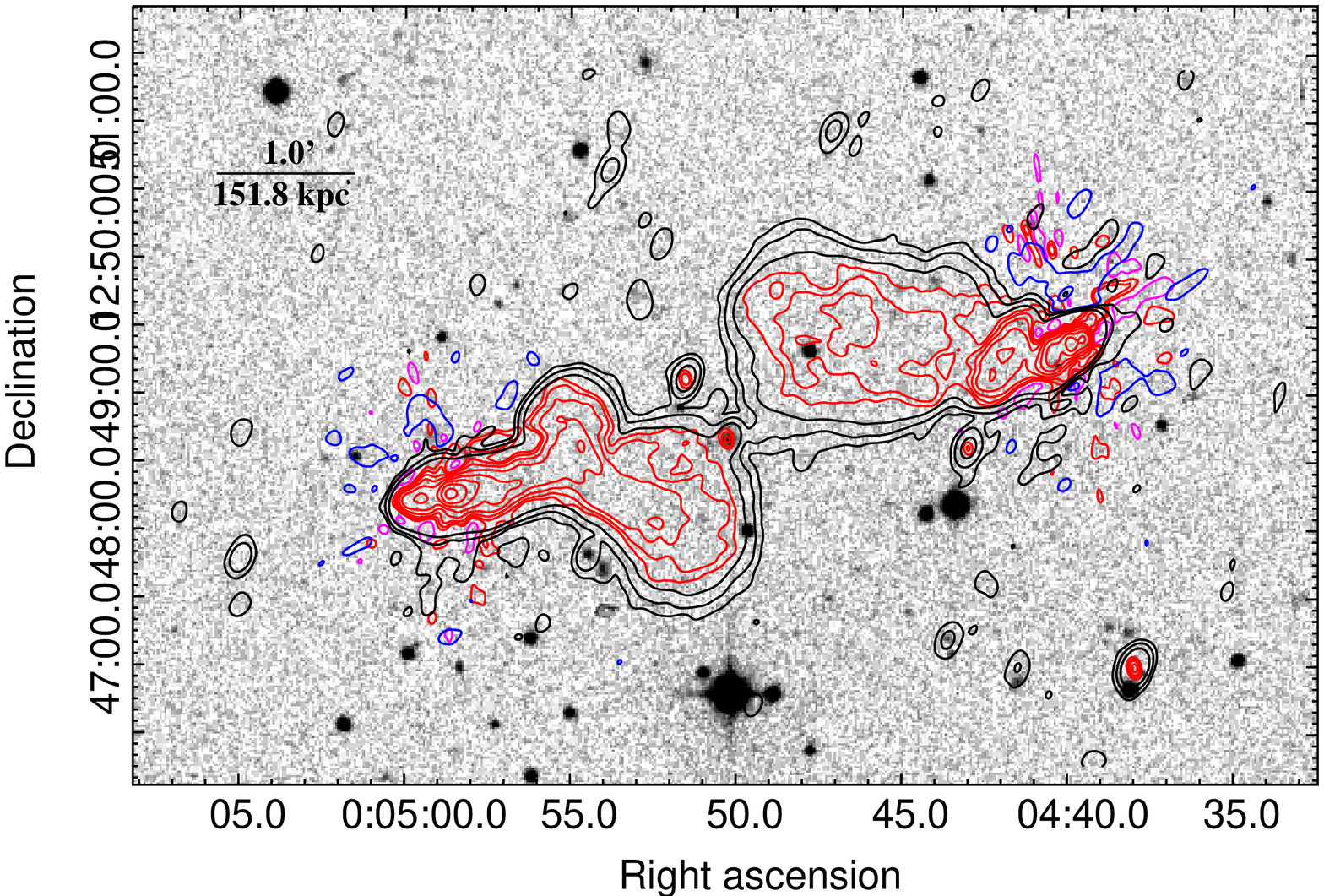} &
\includegraphics[height=6.5cm]{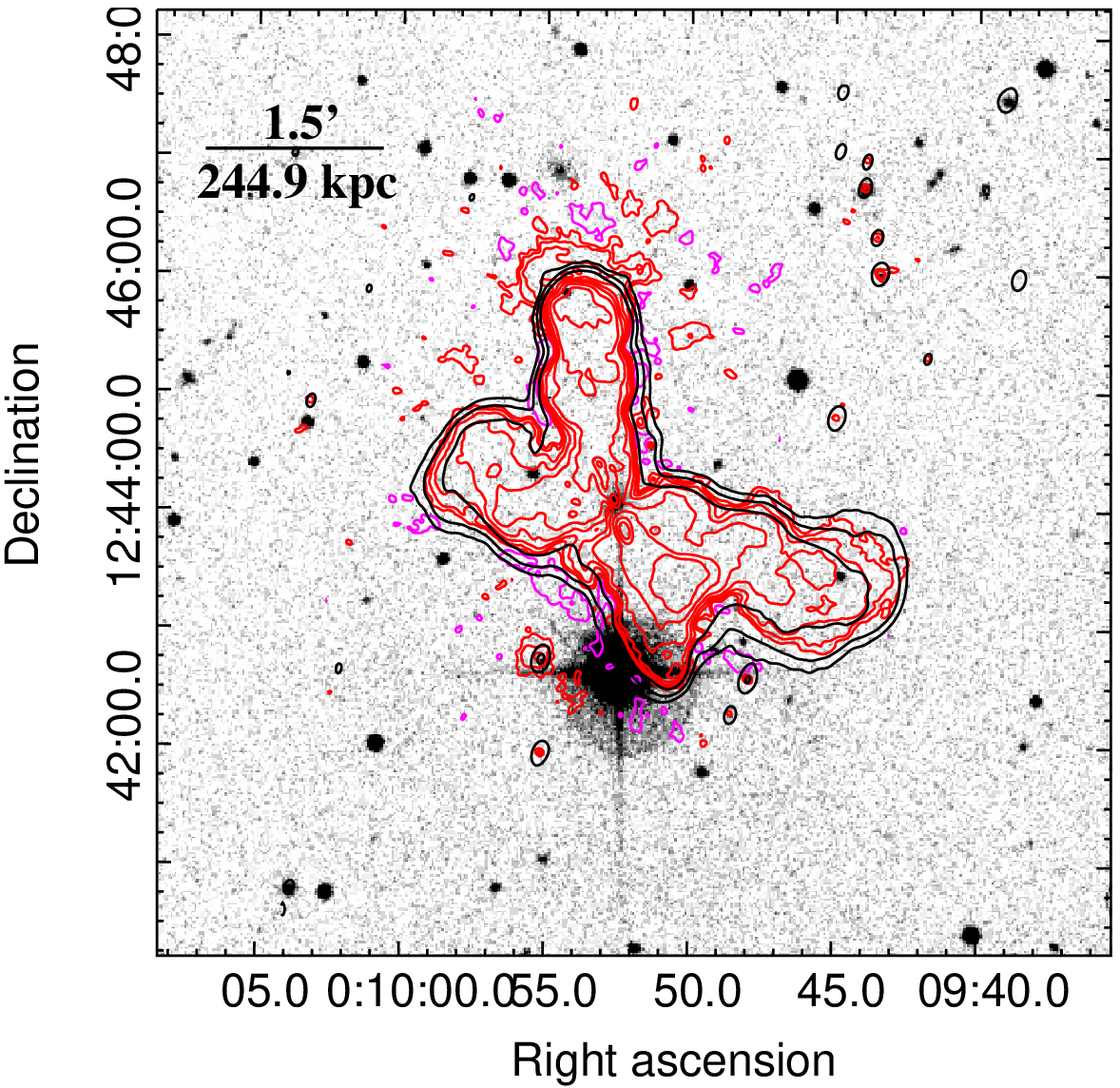} \\
\includegraphics[height=6.8cm]{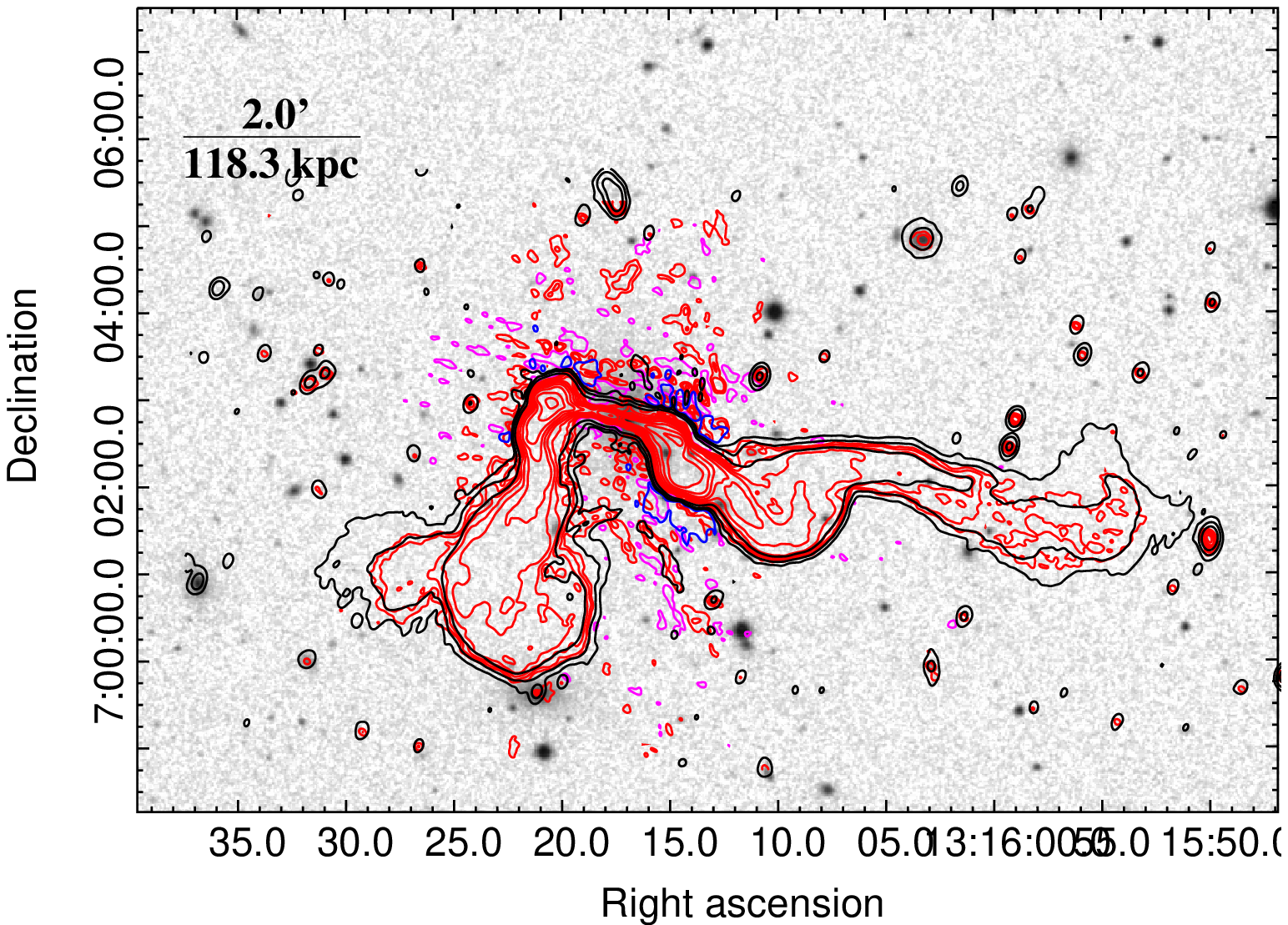} &
\includegraphics[height=6.8cm]{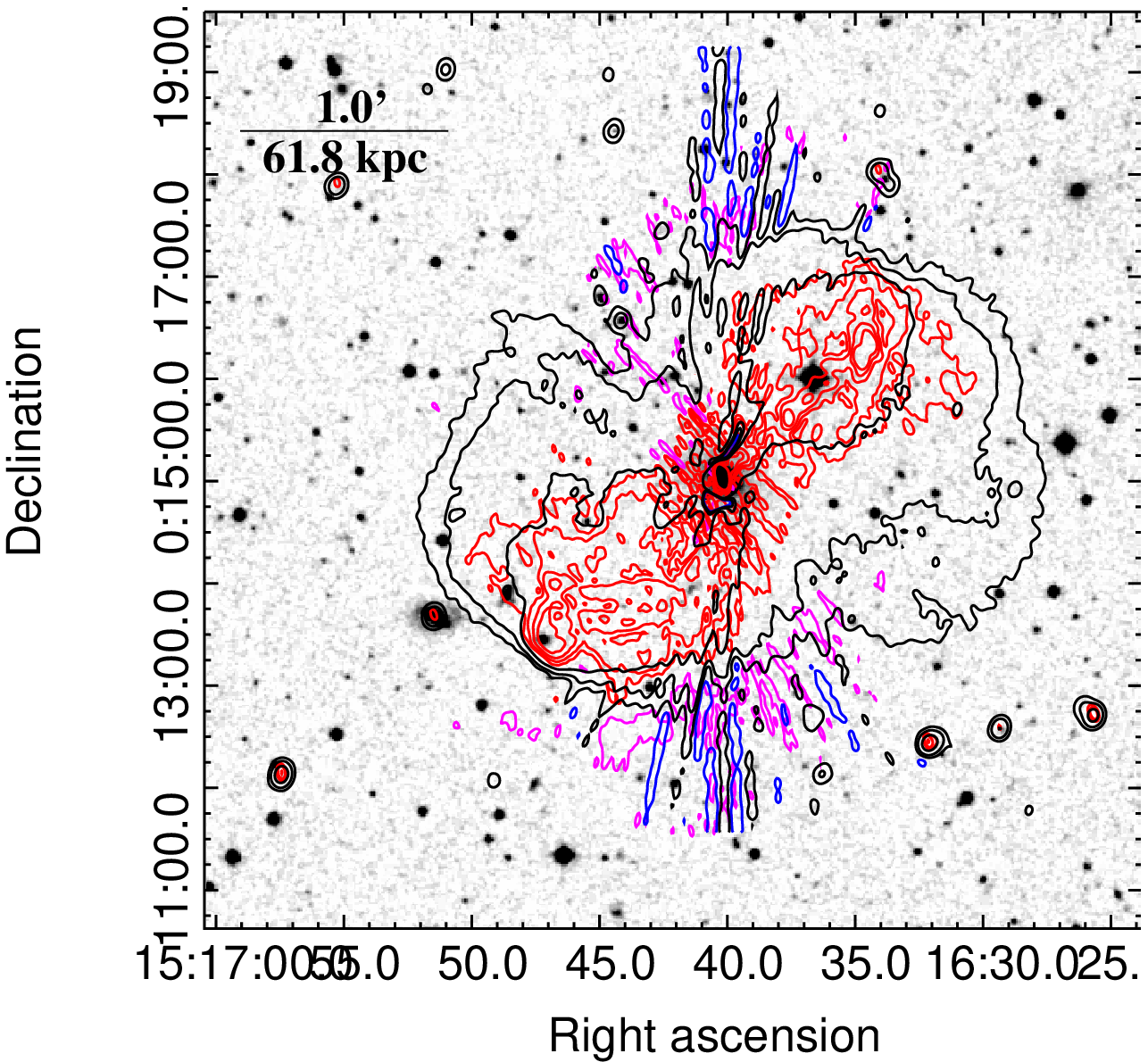} 
\end{tabular}
\caption{The images of our four sample sources, 4C\,12.02 (top-left panel), 4C\,12.03 (top-right panel), CGCG\,044$-$046 (bottom-left panel) and CGCG\,021$-$063 (bottom-right panel).  The uGMRT and MeerKAT images have angular resolutions of $\sim4^{\prime\prime}$ and 7$^{\prime\prime}$, respectively (see also Table~\ref{tab:obs-log}). The radio contours of uGMRT and MeerKAT images are overlaid on the DSS-II (red band) optical images (in gray-scale). The red and black surface brightness contours, with magenta and blue being first negative surface brightness contours, correspond to uGMRT and MeerKAT images, respectively.  The contour levels are \textsc{rms} $\times$ $-$1, 1, 2, 4, etc., and increases by a factor of two. The bar in the top-left corner of each panel image depicts the physical scale for our sample source.}
\label{sample-NVSS}
\end{center}
\end{figure*}

These well-defined selection criteria provided us with a total of 12 sources, 6 FR\,I and 6 FR\,II radio galaxies. We started our project observing the four sources reported in Table~\ref{tab:sample}, which are also members of the GaLactic and Extragalactic All-sky MWA (GLEAM) 4-Jy sample \citep{2020PASA...37...18W,2020PASA...37...17W}.
We note that the morphological classification of CGCG\,021--063 is challenging. The source looks like a FR\,II source in the new MeerKAT and uGMRT observations. This will be addressed in Sec.~\ref{radio-morphology}.
In order to image the sample sources with matching sensitivity,  ($u,v$)-coverage and angular resolution with the two arrays, we observed each target for $\sim$ 3 hours and $\sim$ 3--6 hours respectively with uGMRT and MeerKAT.
The observing logs for the uGMRT and MeerKAT observations are detailed in Table~\ref{tab:obs-log}.

The upgraded GMRT has a hybrid configuration \citep{Guptaetal2017,Swarupetal1991} with half of its 45~m diameter 30 antennas located in a central ($\sim$1~km) compact array and with the remaining antennas distributed in a roughly `Y' shaped configuration giving $\sim$25~km  maximum baseline length.  The antennas in the central square and in a `Y' shaped configuration provide baselines that are comparable to the VLA D-array and B-array configurations, respectively. Hence, a single observation with the uGMRT provides good angular resolution when mapping the detailed source structure with reasonably good sensitivity.
MeerKAT is also a hybrid array of 64 13.5~m diameter dish antennas.
Forty-eight of the 64 interlinked antennas are located in a core region of 1~km in diameter and the other 16 are located outside the core, giving a maximum baseline length of $\sim$8~km \citep{JonasandMeerKAT}.  This configuration of MeerKAT is equivalent to a simultaneous hybrid of VLA B- C- and D-array configurations.
The observations described in this paper were all carried out with the new dual polarisation (RR and LL) 550--850 MHz band (band-4) receivers of uGMRT, and with the dual linear polarisation (horizontal and vertical) 856--1712 MHz (L-band) receivers \citep{Lehmensiek2012,Lehmensiek2014} of MeerKAT.

\begin{figure}
\begin{center}
\begin{tabular}{c}
\includegraphics[height=5.6cm]{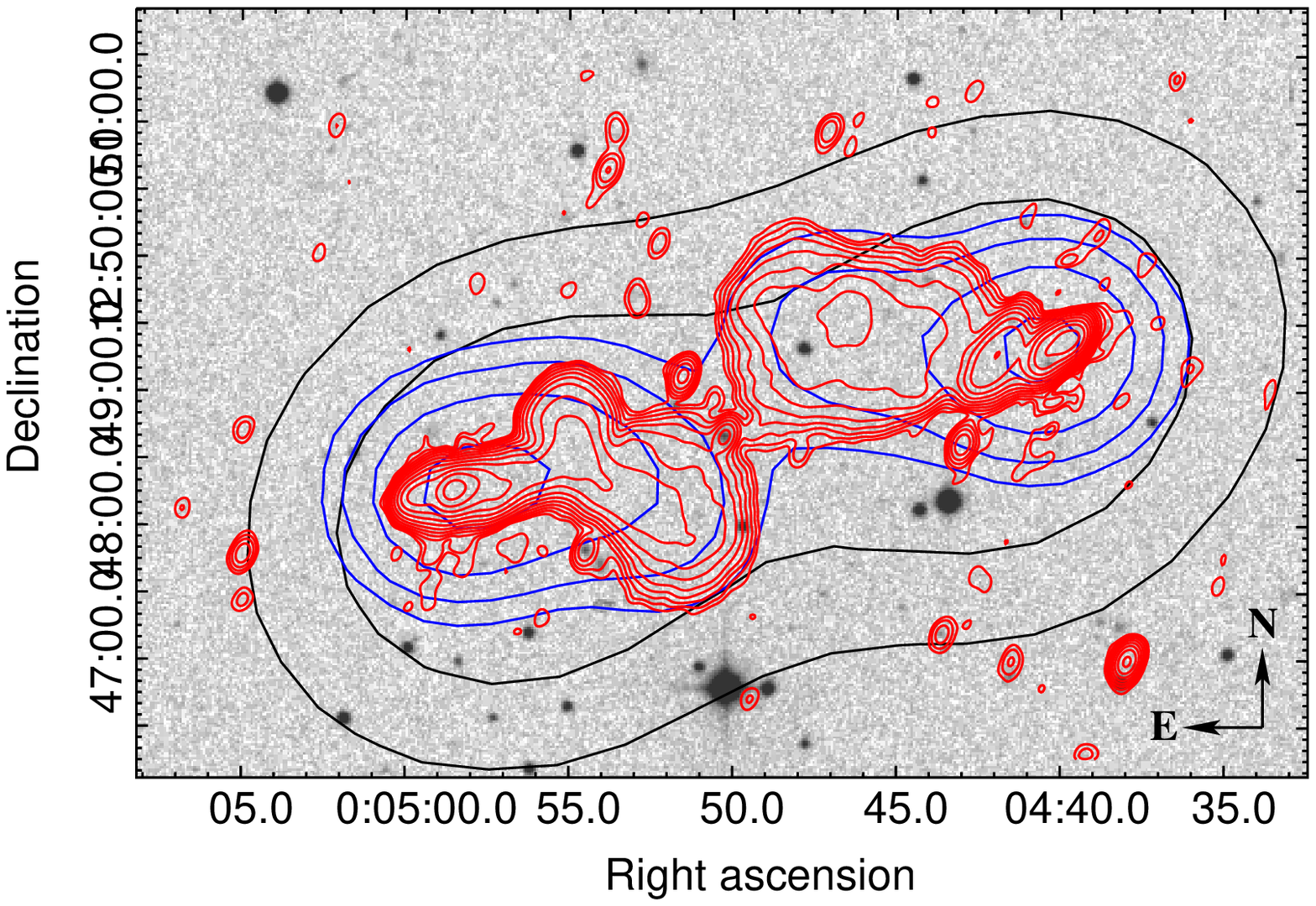}
\end{tabular}
\caption{The image of one of our four sample source, 4C\,12.02 is shown to give an idea of the capabilities of MeerKAT and uGMRT.  The radio contours of the MeerKAT image (red contour), NVSS image (blue contour) and GLEAM image (black contour) are overlaid on the DSS-II (red band) optical image (in gray-scale).  The red, blue, and black surface brightness contours correspond to the contour levels 0.045, 0.090, 0.18, 0.36, 0.72, 1.44, 2.88, 6.0, 12.0, 24.0, 48.0 and 100.0 mJy~beam$^{-1}$, 30, 60, 120 and 240 mJy~beam$^{-1}$, and 800, 1600 mJy~beam$^{-1}$, respectively.  We also show a compass at the bottom right location indicating the north and east directions.  All through our uGMRT and MeerKAT images, we follow this convention.}
\label{fig1}
\end{center}
\end{figure}

\subsection{Data reduction}
\label{data-reduction}

The calibration of the data presented in this paper was carried out following the standard procedures.  However, as the MeerKAT data reduction requires a new approach, we summarize here some of the specific steps that have been performed for both the uGMRT and the MeerKAT in reducing these data.

The uGMRT datasets were calibrated using a standard approach \citep[see][for detailed methodology]{Lal2020}.  Observations of a flux density calibrator at the beginning and at the end of each run were used to correct for flux density scale and bandpass shape; the phase-calibration source was used to correct for phases. The data analysis, in particular editing of bad data, gain and bandpass calibrations, were carried out using the NRAO Astronomical Image Processing System  \textsc{aips}, following standard imaging procedures.
We made an error in providing the observing set-up for the uGMRT observations of CGCG\,044$-$046 source, which had 200 MHz bandwidth with the high spectral resolution, whereas the rest of the sources had 400 MHz bandwidth with a factor of two lower spectral resolution.  Note that this error did not limit us in terms of the angular resolution or the sensitivity (see below).
We used the \citet{PerleyButler} flux density scale using the coefficients in the \textsc{aips} task \textsc{setjy}.
After the above preliminary calibration, the 300-MHz\footnote{The GMRT wideband correlator supports a bandwidth of 400 MHz, 200 MHz, ..., in multiples of 0.5.  The usable bandwidth of band-4 is 300 MHz, from 550 MHz to 850 MHz.} wide dataset was split into six 50~MHz sub-bands for 4C\,12.02, 4C\,12.03 and CGCG\,023$-$061 and four 50 MHz sub-bands for CGCG\,044$-$046.
An area of $\approx$\,1.5$^\circ$\,$\times$ 1.5$^\circ$ was imaged, just bigger than the first null of the uGMRT primary beam in order to correct for antenna-based gains in the direction of each radio source.
A standard self-calibration procedure was performed in \textsc{aips} on each sub-band \citep[see also,][]{LalandRao2007}.
All the calibrated 50~MHz sub-bands data were further stitched together to form full-bandwidth calibrated visibility datasets.  These visibilities were imaged using the \textsc{tclean} task in \textsc{casa}.
A final amplitude-and-phase self-calibration was also carried out to the full-bandwidth calibrated dataset onto the same flux density scale using \textsc{casa}, and the final images were obtained using the task
\textsc{tclean}.
We used 3D imaging (gridder = `widefield') and Briggs weighting (robust = 0.5).

The MeerKAT datasets were also calibrated using a standard approach, but employing some novel software. We used the {\sc CARACal}
pipeline\footnote{\url{https://github.com/caracal-pipeline/caracal}} \citep{caracal1,caracal2} for the initial data reduction. {\sc CARACal} orchestrates standard reduction packages into a single workflow. In this instance, it combined the {\sc Tricolour}\footnote{\url{https://github.com/ska-sa/tricolour}} flagger \citep{tricolour} for radio frequency interference flagging, and standard {\sc CASA} tasks for reference calibration. For the epochs employing B1934$-$638 as the primary calibrator, we used the \citet{PerleyButler} scale to set its flux density scale. For the epoch employing the other standard MeerKAT calibration source, PKS\,0408$-$65, we used a custom component-based field model provided in {\sc CARACal}, converted into model visibilities via the {\sc MeqTrees} package \citep{meqtrees}.   After applying all the reference calibration, the 4096 spectral channels data was averaged down to 1024 spectral channels, and imaged using the {\sc WSCLEAN} package \citep{wsclean}.  We used Briggs weighting (robust = 0), disabled multi-frequency weighting, employed the joined-channel deconvolution and (4th order) polynomial fitting options of {\sc WSCLEAN} to make wideband multi-frequency synthesis images.  We imaged an area of $\approx 2\fdg2\times 2\fdg2$ for the CGCG\,021$-$063 and CGCG\,044$-$046 fields. For the 4C\,12.02 and 4C\,12.03 fields, we  imaged a larger area of $\approx 3\fdg1 \times 2\fdg2$: these sources are sufficiently close that their sidelobes contribute to each other's fields at the sensitivity levels we reach, and thus needed to be deconvolved jointly.  This was followed by a round of phase and delay self-calibration using the {\sc CubiCal}\footnote{\url{https://github.com/ratt-ru/cubical}} package \citep{cubical}.
The 4C\,12.02 field exhibited direction-dependent (DD) effects due to a 0.64 Jy (apparent) off-axis source, which was successfully peeled using {\sc CubiCal}. The other fields did not require DD calibration. The 4C\,12.02 and 4C\,12.03 maps were then slightly improved via a round of amplitude self-calibration using {\sc CubiCal}.  This resulted in noise-limited maps for three of our four sample sources.

The resulting MeerKAT image of CGCG\,021$-$063 retains some radial, north-south oriented artefacts related to the point spread function (\textsc{psf}) centered on the bright core of the source. We were only partially able to mitigate these via self-calibration. These are not due to DD effects, since the core is the dominant source and it is at the centre of the field, nor are they likely to be deconvolution artefacts, as the core is completely unresolved. We hypothesize that they are due to residual nonlinearities in the system response, exacerbated by the effective \textsc{psf} of this observation: since the source is only $15^\prime$ off the equator, the \textsc{psf} exhibits pathologically high sidelobes in the north-south direction, at the 1--2\% level (compared to $<0.1\%$ level in the east-west direction). The dynamic range of $\sim50\,000:1$ achieved here is comparable to that achieved with MeerKAT on other equatorial fields (Ian Heywood, private communication).
Our final images resulted in an angular resolution of $\sim$6$^{\prime\prime}$ for all our target sources.
The residual amplitude errors in each image are of the order of 5\% for uGMRT and 3\% for MeerKAT.
The final 550--850 MHz (band-4) uGMRT and MeerKAT images are shown in Fig.~\ref{sample-NVSS}, \ref{fig1} and \ref{Meer-uGMRT}.

The MeerKAT observations of CGCG\,044$-$046 and CGCG\,021$-$063 included a scan of a bright polarised calibrator, 3C\,286, which allowed us to perform polarisation imaging. Polarisation calibration was done via {\sc CARACal}, using the standard CASA approach. The calibrator B1934$-$638 was used to derive frequency-dependent leakages (\textbf{D}--Jones), while 3C\,286 was used to calibrate cross-hand phase and delay (\textbf{K/X}--Jones). The joined-polarisation mode of {\sc WSClean} was then used to deconvolve Stokes $I, Q, U$ and $V$ maps. The $Q$ and $U$ images of CGCG\,021$-$063 suffered from the same artefacts as the total intensity image, so we did not use them in our further analysis.
We include here results from our polarisation observations using MeerKAT for CGCG\,044$-$046 (see Sec.~\ref{sec.sum-conc}).

\section{The images}
\label{sec.morph-spec}

\begin{figure*}
\begin{center}
\begin{tabular}{rr}
\includegraphics[width=8.55cm]{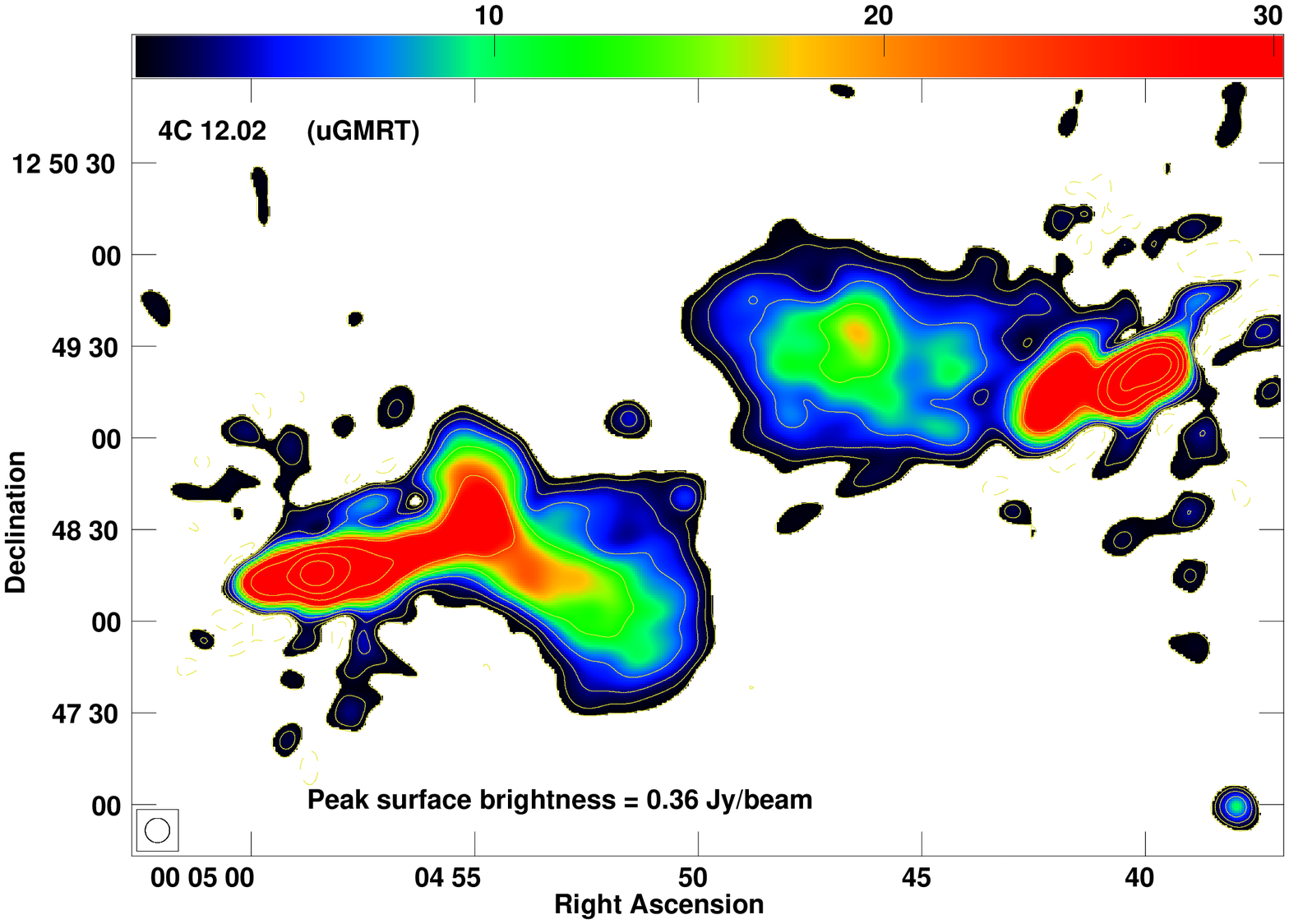} &
\includegraphics[width=8.55cm]{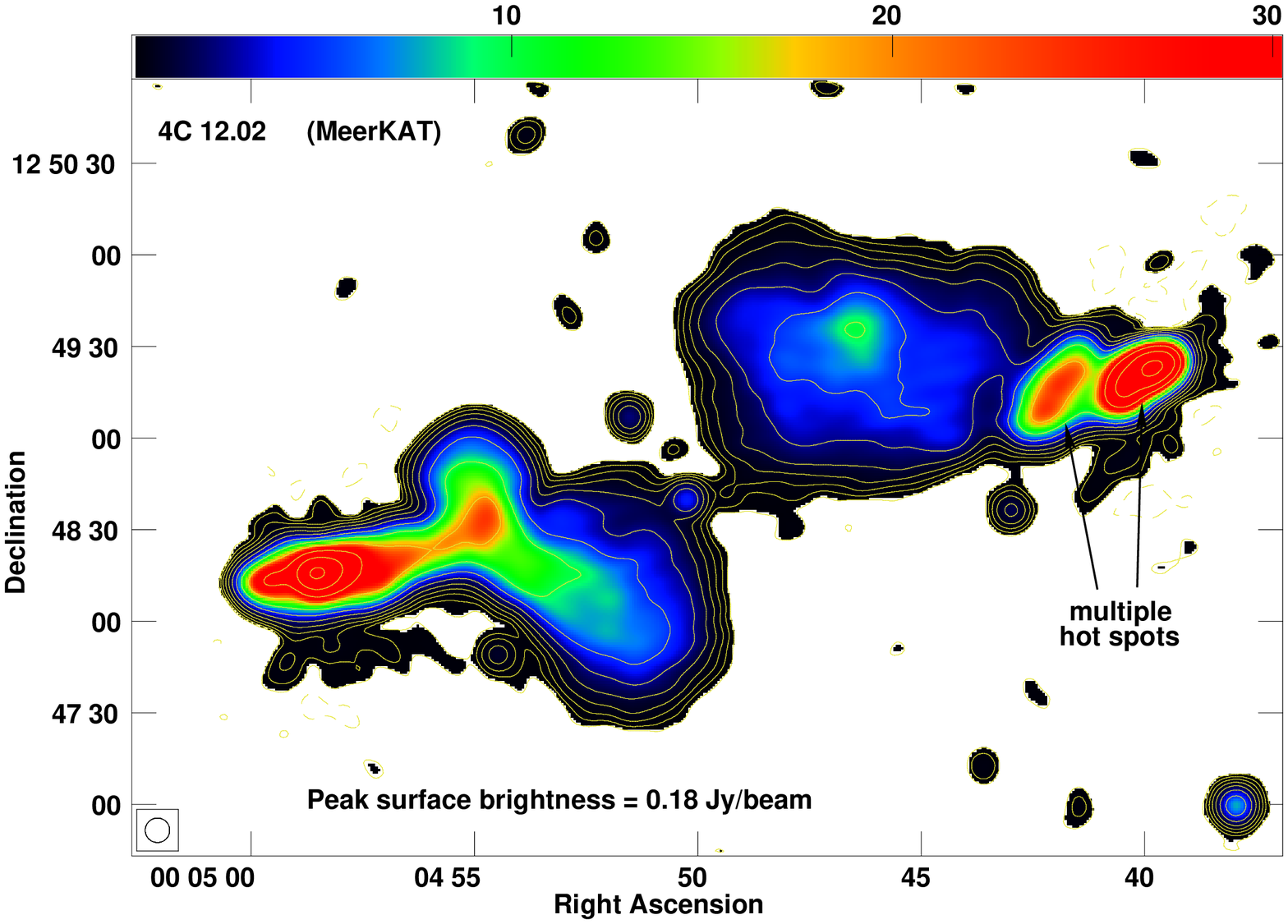} \\
\includegraphics[width=8.55cm]{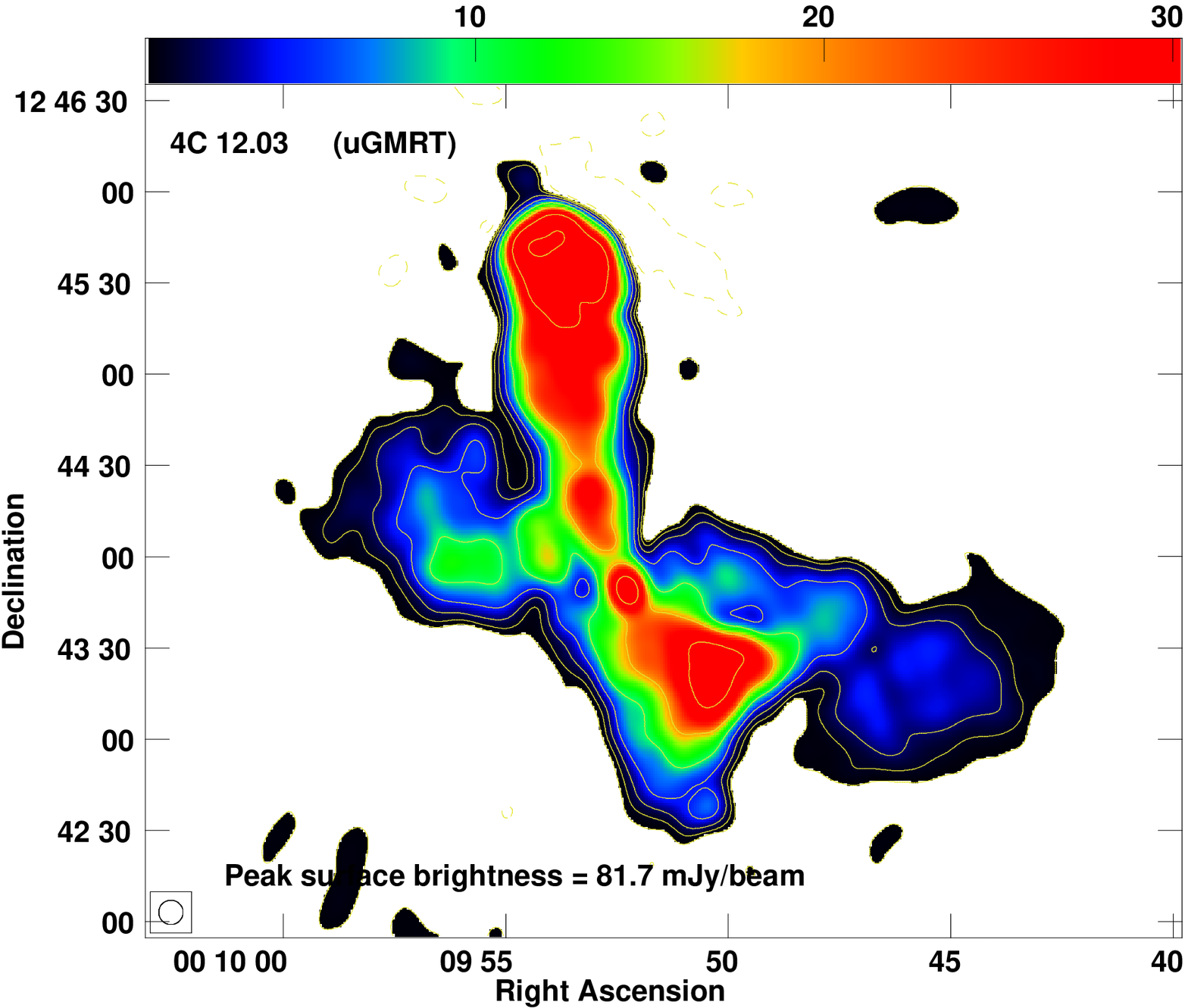} &
\includegraphics[width=8.55cm]{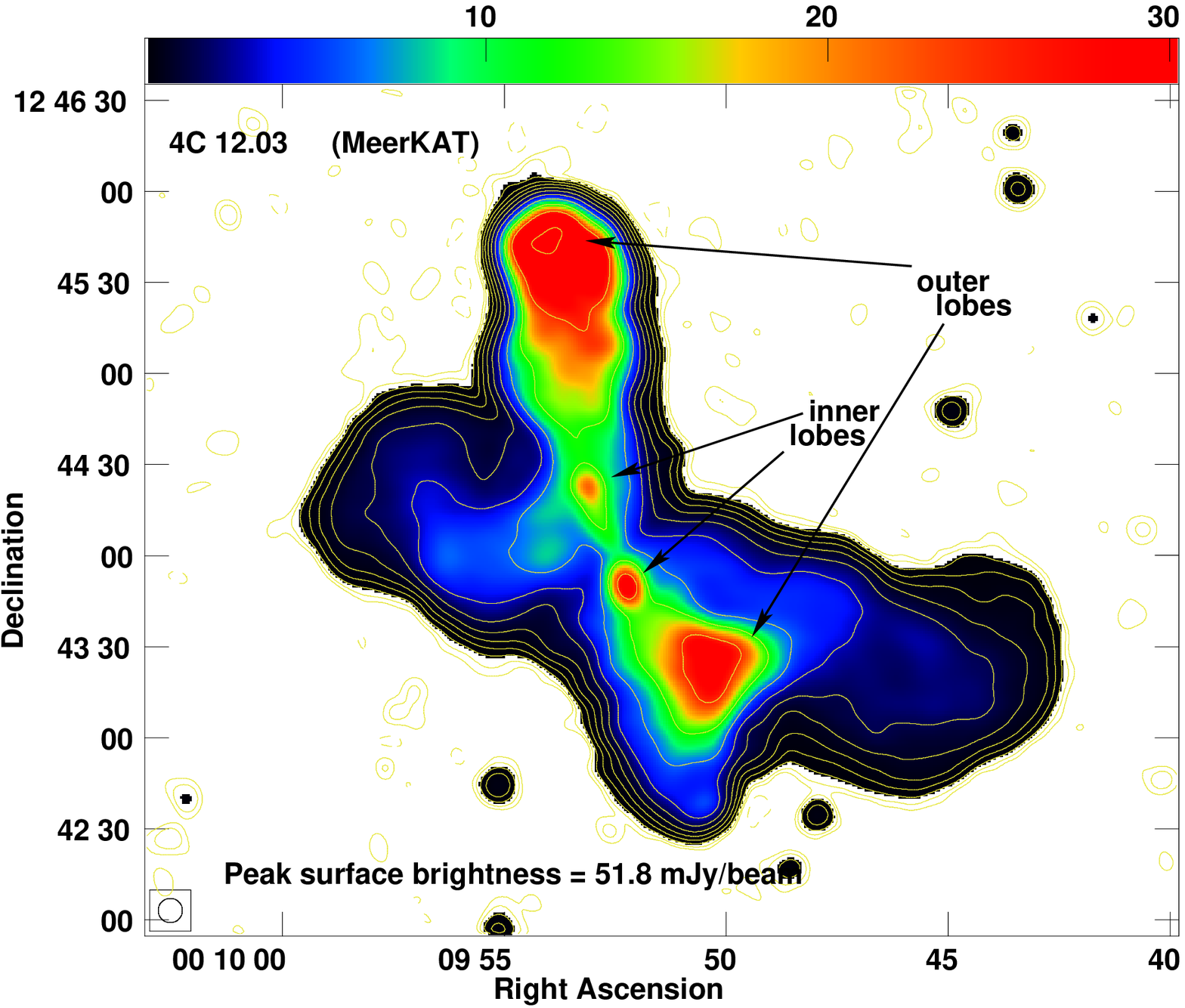} \\
\includegraphics[width=8.55cm]{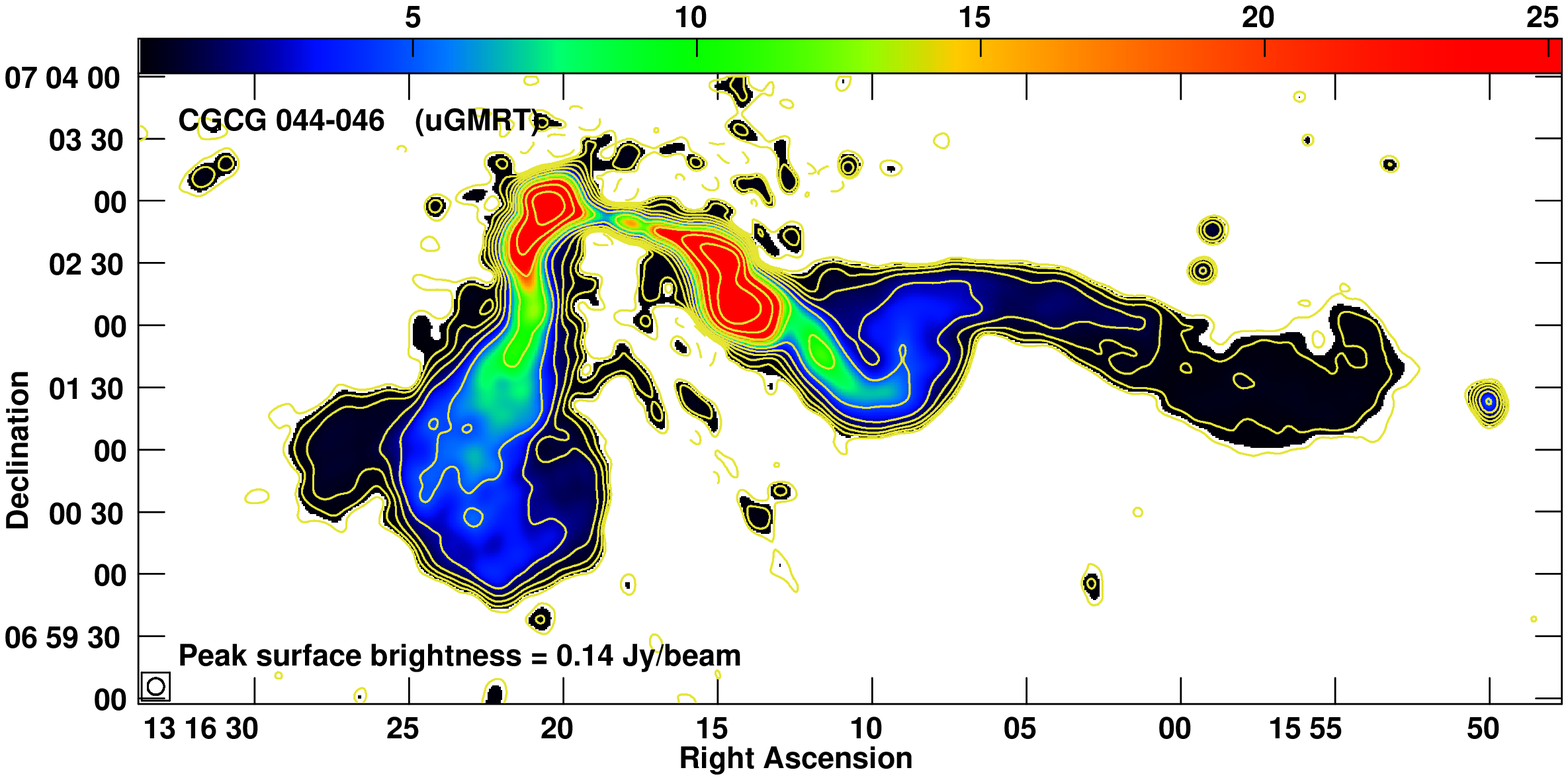} &
\includegraphics[width=8.55cm]{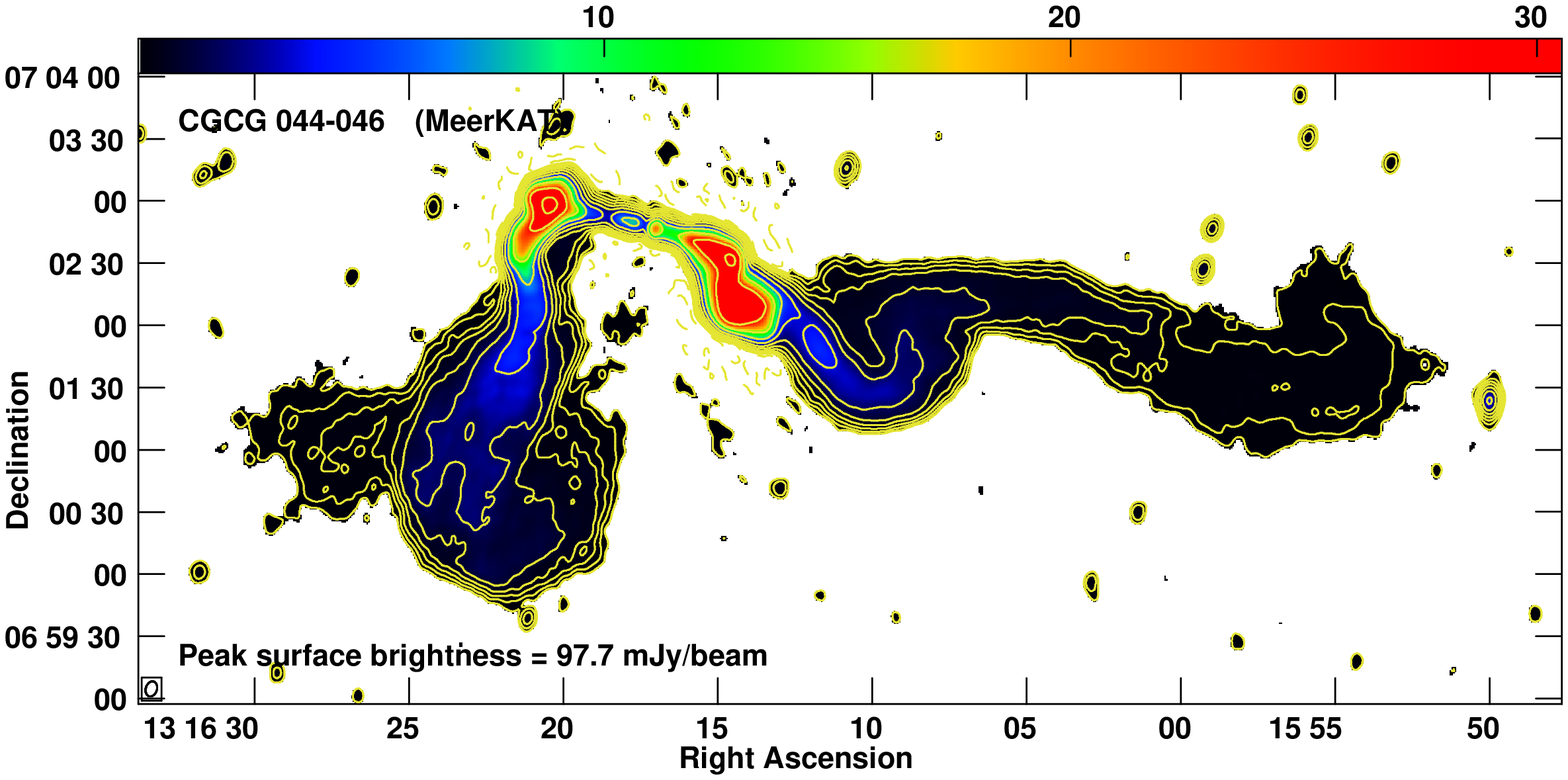}
\end{tabular}
\caption{Images of four sample sources at band-4 using the uGMRT (left panel) and at L-band using the MeerKAT (right panel).
These radio sources are in the order presented in Table~\ref{tab:sample}.
The lowest radio contour plotted is three times the local \textsc{rms} noise and increasing by factors of 2.  The local \textsc{rms} noise and the beam size (along with the position angle) are denoted in Table~\ref{tab:obs-log}.}
\label{Meer-uGMRT}
\end{center}
\end{figure*}

\setcounter{figure}{2}
\begin{figure*}
\begin{center}
\begin{tabular}{rr}
\includegraphics[width=8.55cm]{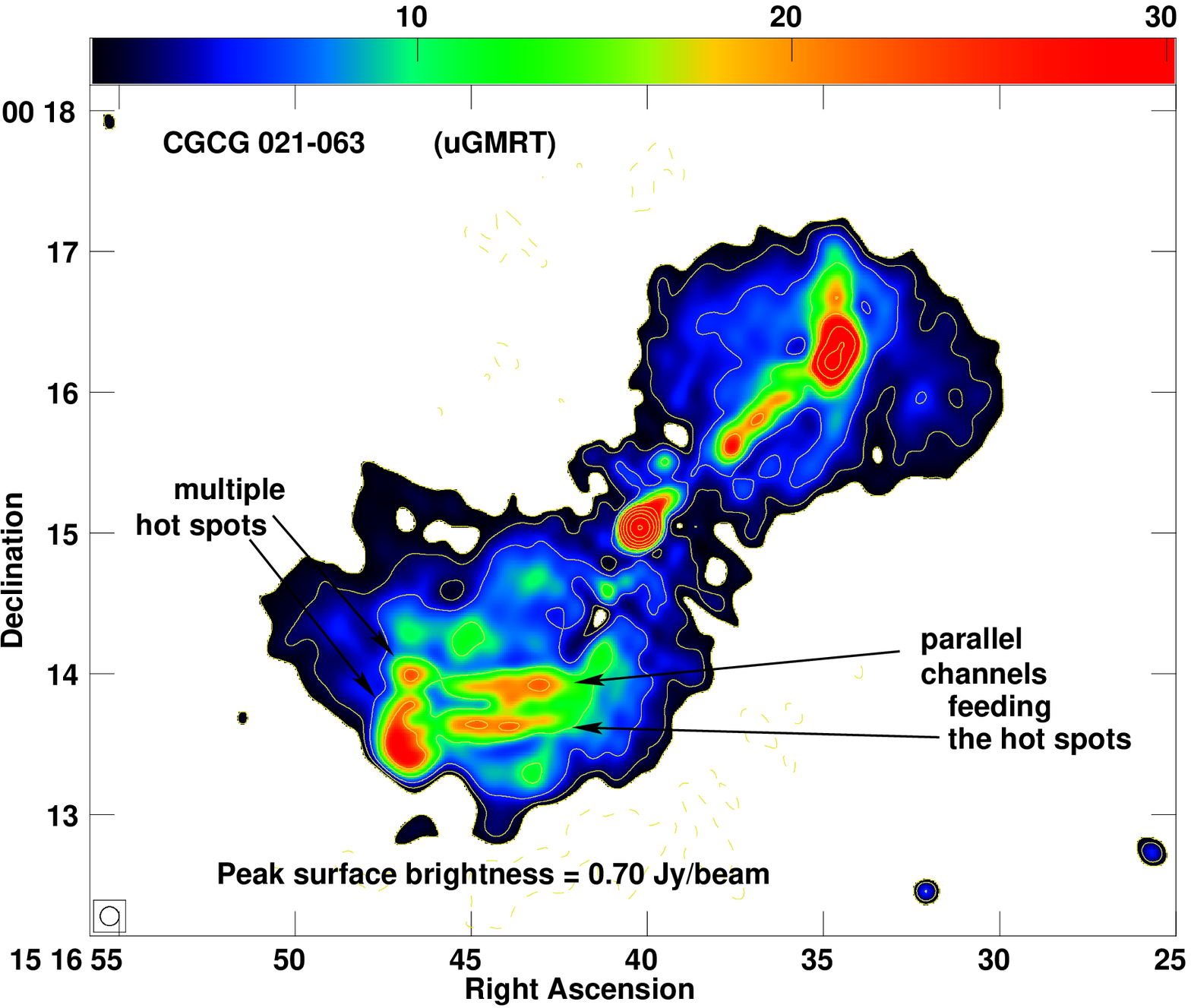} &
\includegraphics[width=8.55cm]{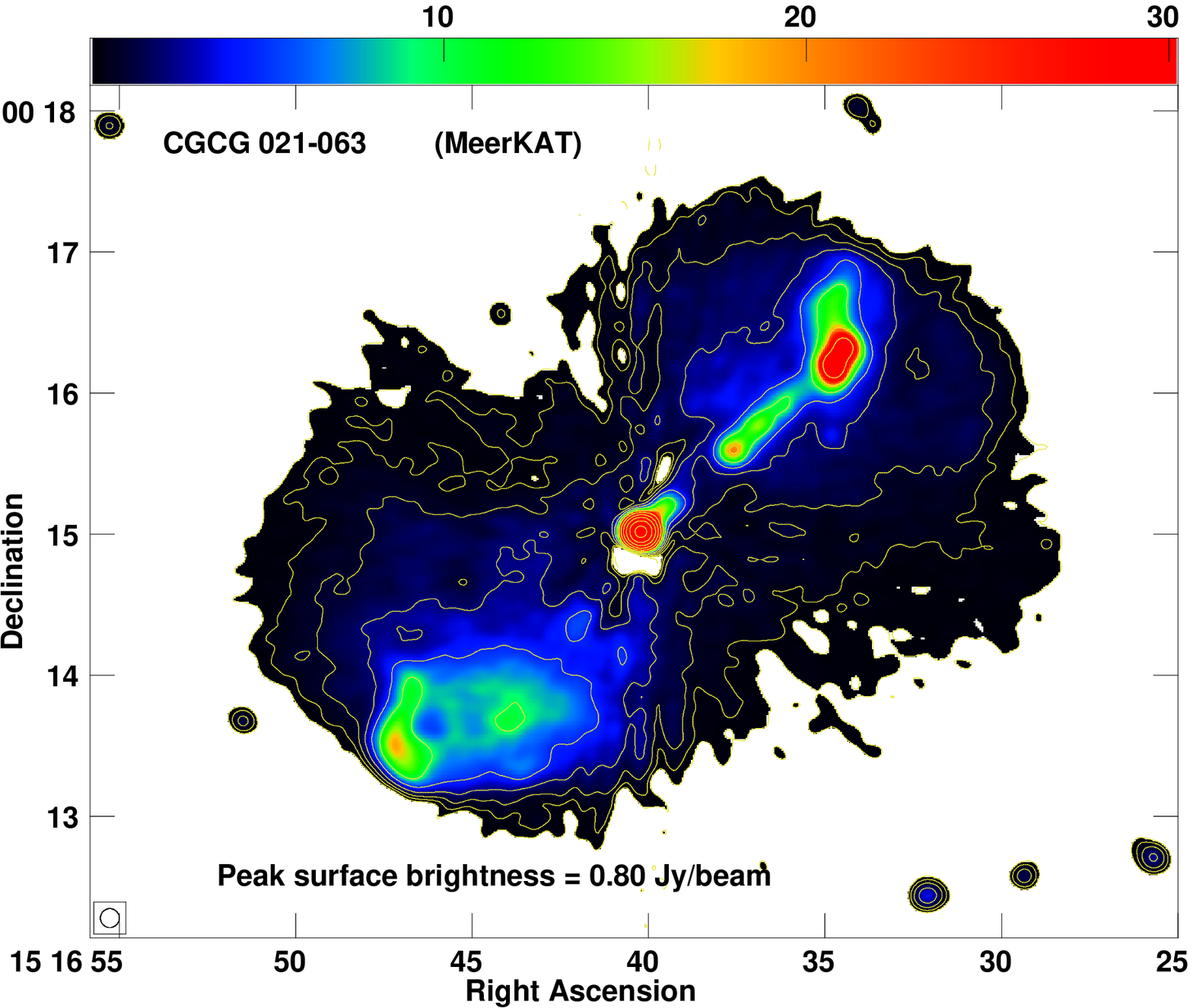}
\end{tabular}
\caption{Continued.}
\end{center}
\end{figure*}

Our uGMRT and MeerKAT observations revealed several new sources in the fields of view in addition to deep images of our four targets sources.
Fig. \ref{sample-NVSS} shows the uGMRT and MeerKAT contours overlaid on the DSS-II (red band) optical image.
Fig.~\ref{fig1} is an image of 4C\,12.02 showing the MeerKAT radio contours together with radio contours of NVSS and GLEAM images overlaid on the DSS-II (red band) optical image, to faithfully present the capabilities of MeerKAT and uGMRT.  The compass depicts north and east and all through we follow this convention.
Fig. ~\ref{Meer-uGMRT} provides colour scale with contour levels overlaid for each dataset (left and right panels showing the uGMRT and MeerKAT image respectively).
It clearly shows that in almost all cases, barring CGCG\,044$-$046, the surface brightness declines sharply at the edges of the visible radio lobes and of the low surface brightness features. We thus conclude that we have not missed any part of the source that fades into the noise. Details on each source are provided below.

\subsection{Radio morphology}
\label{radio-morphology}

\subsubsection{4C\,12.02} It is an FR\,II radio galaxy whose optical counterpart (a galaxy) is located at redshift $z$ = 0.143. 
\citet{1975MmRAS..79....1S} first noted it as a triple source, with a radio core, and two radio lobes located on opposite sides along the east-west direction.  Recently, \citet{Xiaolongetal} reported it as a representative giant X-shaped radio source based on the GMRT TGSS\_ADR1 \citep{Intemaetal}.  

Our uGMRT and MeerKAT images are shown in Fig.~\ref{sample-NVSS} (top-left panel) and Fig.~\ref{Meer-uGMRT} (top-left and top-right panels respectively). The angular resolution of the uGMRT provides important insight into the morphology of the hot spots, while the MeerKAT image is better suited to highlight the details and extent of the lobes. 
The morphology of the back-flows from the lobes is similar to what has been found for PKS\,2014$-$55 \citep{Cottonetal}.
Both hot spots are resolved with multiple peaks. There are at least two hot spots near the termination of the west jet and probably a string of them in the east one.
Both the east and west lobes show substructure.
The north-western lobe flares abruptly at the peak of the hot spot closer to the core, while the southern lobe expands perpendicular to the main axis of the radio galaxy.
In the west lobe, the hot spot before the bow shock is clearly extended in an arc perpendicular to the source major axis, which may suggest some sort of reflected, internal shock.
The hot spots in the east and west lobes are not exactly co-linear if the line goes through what appears to be the active galactic nucleus (AGN). It is not clear that a projection effect could do this.
The total flux density of the source is S$_{\rm 0.69~GHz}=3.50~\pm$0.19 Jy and 
S$_{\rm 1.28~GHz}=2.15~\pm$0.07 Jy.
The angular extent of the source is $\sim5^{\prime}$, corresponding to $\sim$760 kpc, {\it i.e.} a giant radio galaxy \citep[see][]{2016MNRAS.461.3165B}. 

\subsubsection{4C\,12.03} This X-shaped radio source is associated with an elliptical host galaxy \citep{Heckmanetal}, classified as a low emission line radio galaxy located at redshift $z$ = 0.156 \citep{1983MNRAS.204..151L}.  
\citet{LalandRao2007} reported that 240 MHz and 610 MHz GMRT images show a symmetrical structure whose extent is $\sim 4^{\prime}$ along both axes, corresponding to $\sim$650 kpc.
The overall X-shaped morphology has been explained by a million-year precession period \citep{2011ApJ...734L..32G}, or as a rapid realignment of a central supermassive black hole accretion disk system due to a relatively recent merger of a supermassive binary black hole \citep{LalandRao2007,MerrittEkers} or as a result of disk instability \citep{Dennett2002}.

Our new uGMRT and MeerKAT images, shown in Fig.~\ref{sample-NVSS} (bottom-left panel) and Fig.~\ref{Meer-uGMRT} (middle-left and middle-right panels respectively), confirm the X-shaped radio morphology.
The northern and southern jets leading to the north and south hot spots respectively, form the active axis, whereas the east-west axis forms the low-surface brightness wings.
The overall north-south morphology of the radio galaxy seems to follow an arc, and shows a clear asymmetry:  the northern component culminates in a clear hot spot, as in FR\,IIs. On the other hand, the southern component could be either FR\,I or FR\,II.
Both the western and eastern low-surface brightness wings are suggestive of a hydrodynamic back-flow as seen in PKS\,2014--55 \citep{Cottonetal}. 

The angular resolution of both our images clearly shows the presence of two 
inner brightness peaks (labelled as "inner lobes" in Fig.~\ref{Meer-uGMRT}, middle-right panel), forming an inner double structure, whose total extent is $\sim$ 80 kpc. These two features are perfectly aligned with the large north-south axis \citep[see also][]{Schoenmakersetal}.  Thus, the structures of the source possibly place it in the category of restarted sources.
The overall extent of this radio galaxy is comparable to the earlier images, but broader extended emission is detected in the east-west radio lobes, whose outer parts show a filamentary structure at both frequencies.
The total flux density of the source is S$_{\rm 0.69~GHz}=3.14~\pm$0.18 Jy and S$_{\rm 1.28~GHz}=2.06~\pm$0.06 Jy.

\subsubsection{CGCG\,044$-$046} The source is identified as a m = 14.2 cD galaxy and is associated with the Zwicky cluster 1313.7$+$0721 at $z$ = 0.050145 \citep{ZwickyHerzog}. Indeed the bent radio morphology is typical of radio galaxies at the centre of galaxy clusters and groups. Some notable examples are ESO~137--006 \citep{Ramatsokuetal} and 3C\,465 \citep{Eileketal}, but see also \citet{ODonoghueetal1990}, and \citet{Garonetal} for a recent analysis.

The radio source was first studied in detail by \citet{Patnaik1984} and \citet{Patnaik1986}, who noticed the asymmetry between the sharp bend of about 90$^{\circ}$ in the eastern tail and the more gentle bend in the western one, with diffuse emission further out in both.  This asymmetry is confirmed in both our images (see Fig.~\ref{sample-NVSS}, upper-right panel, and Fig.~\ref{Meer-uGMRT}, third row, left and right panels for uGMRT and MeerKAT respectively), whose sensitivity allows the detection of emission from the radio tails to much larger distances than previously detected at these frequencies (see also Sec.~\ref{cgcg044}).  The morphology suggests that it possibly belongs to the wide-angle tail class of radio galaxies.  The extension of the western tail was also detected at lower frequencies via GLEAM though at much lower angular resolution \citep{2020PASA...37...18W,gleametal}.

\begin{table*}
\begin{center}
\caption{The total intensity and spectral index for our sample sources. The integrated flux densities quoted are in Jy along with corresponding error-bars when available.}
\label{tab:in-fd-spec}
\begin{tabular}{lrrrrrccc}
\hline
\multicolumn{1}{c}{Source} & $S_{\rm 160\,MHz}$ & $S_{\rm 318\,MHz}$ & $S_{\rm 408\,MHz}$ & $S_{\rm 690\,MHz}$ & $S_{\rm 1284\,MHz}$ & $S_{\rm 1400\,MHz}$ & $S_{\rm 4850\,MHz}$ & $S_{\rm 15000\,MHz}$ \\
  & \multicolumn{1}{c}{(Jy)} & \multicolumn{1}{c}{(Jy)} & \multicolumn{1}{c}{(Jy)} & \multicolumn{1}{c}{(Jy)} & \multicolumn{1}{c}{(Jy)} & \multicolumn{1}{c}{(Jy)} & \multicolumn{1}{c}{(Jy)} & \multicolumn{1}{c}{(Jy)} \\
\multicolumn{1}{c}{(1)} & \multicolumn{1}{c}{(2)} & \multicolumn{1}{c}{(3)} & \multicolumn{1}{c}{(4)} & \multicolumn{1}{c}{(5)} & \multicolumn{1}{c}{(6)} & \multicolumn{1}{c}{(7)} & \multicolumn{1}{c}{(8)} & \multicolumn{1}{c}{(9)} \\
\hline\noalign{\smallskip}
4C\,12.02     & 9.1 $\pm$0.82$^{\rm a}$&                          &                         & 3.50 $\pm$0.19 & 2.15 $\pm$0.07 & 1.88 $\pm$0.28$^{\rm e}$ & 0.46 $\pm$0.07$^{\rm g}$ &                 \\
4C\,12.03     & 7.6 $\pm$0.95$^{\rm b}$&                          & 4.45 $\pm$0.20$^{\rm d}$& 3.14 $\pm$0.18 & 2.06 $\pm$0.06 & 1.90 $\pm$0.29$^{\rm e}$ & 0.54 $\pm$0.07$^{\rm g}$ &                 \\
CGCG044$-$046 & 5.6 $\pm$0.84$^{\rm a}$&                          &                         & 3.24 $\pm$0.16 & 2.05 $\pm$0.07 & 1.93 $\pm$0.07$^{\rm f}$ & 0.97 $\pm$0.15$^{\rm h}$ &                 \\
CGCG021$-$063 & 9.1 $\pm$0.82$^{\rm a}$& 5.74 $\pm$0.26$^{\rm c}$ &                         & 3.77 $\pm$0.20 & 2.80 $\pm$0.09 & 2.69 $\pm$0.10$^{\rm f}$ & 1.64 $\pm$0.23$^{\rm i}$ & 1.06 $\pm$0.01$^{\rm j}$ \\
 \hline
\end{tabular}
\end{center}
\begin{flushleft}
The flux densities of four radio sources at 550-850~MHz band (Col. 5) and at L-band (Col. 6) are reported.
The flux density is determined using \textsc{aips} task \textsc{tvstat} for irregular shaped radio sources.  The error-bar on the measurement of flux density is based on the \textsc{rms} noise as evaluated in the irregular shaped polygon.
The rest of the measurements at 151 MHz (Col. 2), 1400 MHz (Col. 6) and 4850 MHz (Col. 7) are from the NASA Extragalactic Database (NED). The spectral index (Col. 8) corresponds to the line representing the best fitting regression to this data. \\
      References: The references for flux density measurements from NED are as follows: (a) \citet{1995AuJPh..48..143S}; (b) \citet{1967MmRAS..71...49G} and the measurement is at 178~MHz; (c) \citet{1981A&AS...45..367K}; (d) \citet{1981MNRAS.194..693L}; (e) \citet{1990PKS90.C...0000W}; (f) \citet{1992ApJS...79..331W}; (g) \citet{1991ApJS...75.1011G}; (h) \citet{1991ApJS...75....1B}; (i) \citet{1995ApJS...97..347G}; (j) \citet{2014MNRAS.438.3058R}.
\end{flushleft}
\end{table*}

The radio galaxy shows an inner pair of straight jets, similar to FR\,I jets, which culminate in two high surface brightness knots at $\sim$ 50 kpc from the core. Beyond this region, the eastern jet flares to form a tail, which bends sharply by $\sim$90$^{\circ}$, heading toward the south, consistent with \citet{Patnaik1984}. 
Our images clearly show another bend at the southern end of the tail, behind the tail itself, as is clear from the radio contours (see Fig.~\ref{Meer-uGMRT}) and from the spectral analysis (see Sec.~\ref{spectral-analysis}).
The transition between the western jet and the tail is quite sharp. The tail shows very little transverse expansion and makes a few wiggles before fading. Both our images reveal filamentary structure in the tails, whose projected length is about 150 and 300~kpc respectively for the eastern and the western tails, as measured using the MeerKAT data.
Such sharp bends cannot be explained by projection effects, and are suggestive of reflection or refraction at discontinuities. 
The total flux density of the source is S$_{\rm 0.69~GHz}=3.24~\pm$0.16 Jy and S$_{\rm 1.28~GHz}=2.05~\pm$0.07 Jy.

\subsubsection{CGCG\,021$-0$63} Very little information is reported in the literature on the arcsecond-scale properties of this radio galaxy, characterised by a compact, {\it i.e.} $<$0.2 milli-arcsec in size \citep{Dodsonetal}, radio core.
Using the FIRST survey image, \cite{Proctor} reported that this source has a resolved compact component, with triple extended structure and possibly a $S$ or $Z$-shaped structure.  
Our uGMRT and MeerKAT images (Fig.~\ref{sample-NVSS}, bottom-right panel, and bottom row in Fig.~\ref{Meer-uGMRT}, left panel and right panel respectively) are dynamic range limited, most likely because of the strong compact core. However, they clearly show that the radio emission of CGCG\,021--063 has two components: a radio galaxy, whose projected linear size is $\sim 280$ kpc that is embedded in a low-surface brightness cocoon of radio emission. 
The inner radio galaxy and the lobes have an FR\,II morphology, with only one visible jet and two hot spots of moderate brightness. It is interesting to note that the west jet seems to propagate beyond the hot spot or working surface of the jet.
There is filamentary structure in the east lobe which is well reproduced in both images and there are also clearly beads in the west jet. We interpret the cocoon of fainter emission embedding the radio galaxy as the result of the back-flow from both jets. This emission is fairly symmetrical, suggesting that there is little relative movement of the galaxy through the intra-galactic medium (IGM).

No information on the local environment of the optical host is available in the literature.
The total flux density of the source is S$_{\rm 0.69~GHz}=3.77~\pm$0.20 Jy and S$_{\rm 1.28~GHz}=2.80~\pm$0.09 Jy.

\section{Spectral analysis}
\label{spectral-analysis}

The spectral index analysis is a useful tool to investigate the life cycle of radio sources and to better understand their radio morphologies.
In the following we present the integrated radio spectra for all sources. We further show uGMRT-MeerKAT spectral index imaging for CGCG\,044--046 and in-band MeerKAT spectral imaging for all sources.

\subsection{Integrated radio spectra}
\label{rad-spec}

\begin{figure}
\begin{center}
\begin{tabular}{c}
\includegraphics[height=6.2cm]{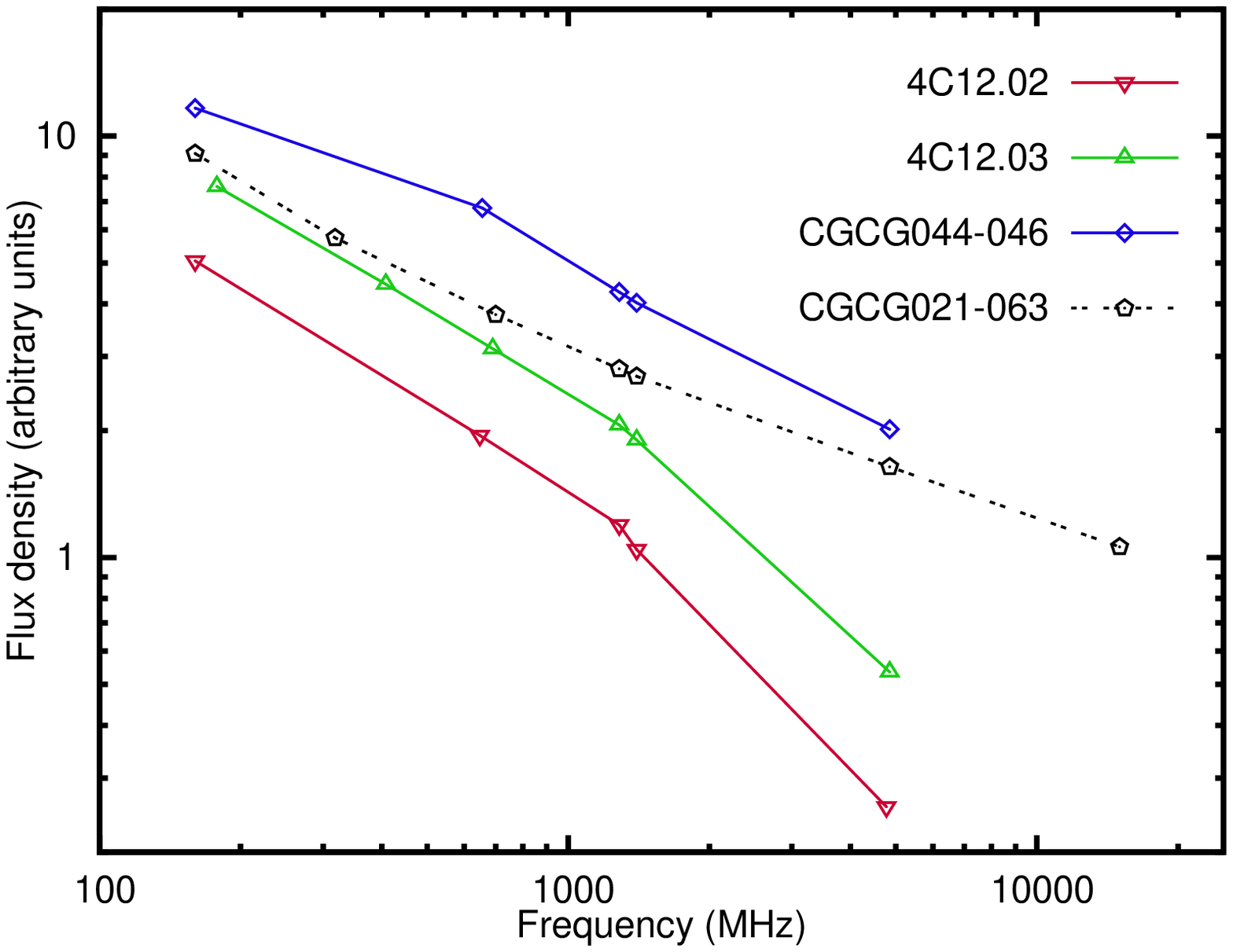}
\end{tabular}
\caption{Integrated flux densities of our sample radio sources at 550--850 MHz band of the uGMRT and 856--1712~MHz band of MeerKAT;
data at other frequencies are from the literature (see Table~\ref{tab:in-fd-spec} for references). The error-bars, not plotted are less than two times the size of the symbols. The spectra (and the data-points) are shifted with respect to one another for clarity.}
\label{fd-spectra}
\end{center}
\end{figure}

We complemented our flux density measurements with literature information.
Table~\ref{tab:in-fd-spec} (Cols. 2--9) lists the integrated flux density we collected for our sources over a broad range of frequencies along with error-bars. 
The corresponding integrated spectra are shown in Fig. \ref{fd-spectra}. 

Our data points are very well aligned with the literature data, which confirms the reliable calibration for both uGMRT and MeerKAT observations.
Note that we have used an identical polygon for our uGMRT and MeerKAT images to determine integrated flux density, making sure that we do not include the obvious artefacts in the images.
Inspection of Fig. \ref{fd-spectra} clearly shows that the spectra of our sources have slightly different behaviours.
Those of CGCG\,044--046 and CGCG\,021--063 are reasonably well-fitted by a single power law, with $\alpha = -$0.52 $\pm$0.09 and $-$0.44 $\pm$0.11, respectively. Both these values are very flat, which is remarkable if we consider that the integrated flux density measurements include the diffuse emission tails and radio lobes, whose spectra are typically steeper, {\it i.e.} $\alpha\simeq -$0.8. This suggests that for both sources the active components, {\it i.e.} the core and inner jets, are the dominant source of emission over a broad range of frequencies.
The spectra of 4C\,12.02 and 4C\,12.03 on the other hand, show a clear break at $\sim$ 1.2 GHz. Further, for both sources the spectra have $\alpha \sim -$0.7 down to 1.284 MHz, which steepens to $\alpha \sim -$1.2 at higher frequencies, suggesting the dominant role of the radio lobes at low frequency, as qualitatively suggested by our total intensity images in both cases.

\begin{figure*}
\begin{center}
\begin{tabular}{rr}
\includegraphics[width=7.4cm]{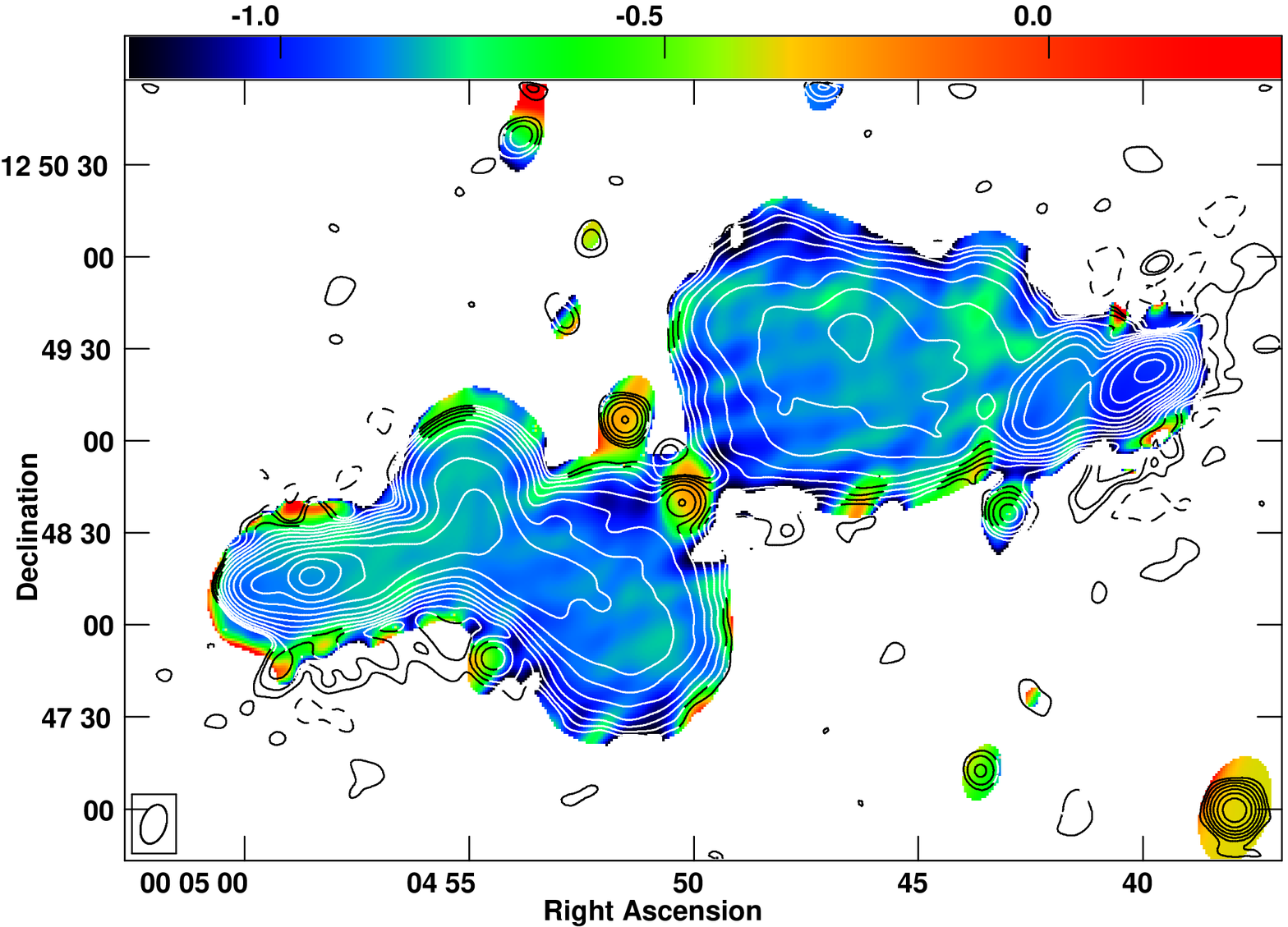} &
\includegraphics[width=6.2cm]{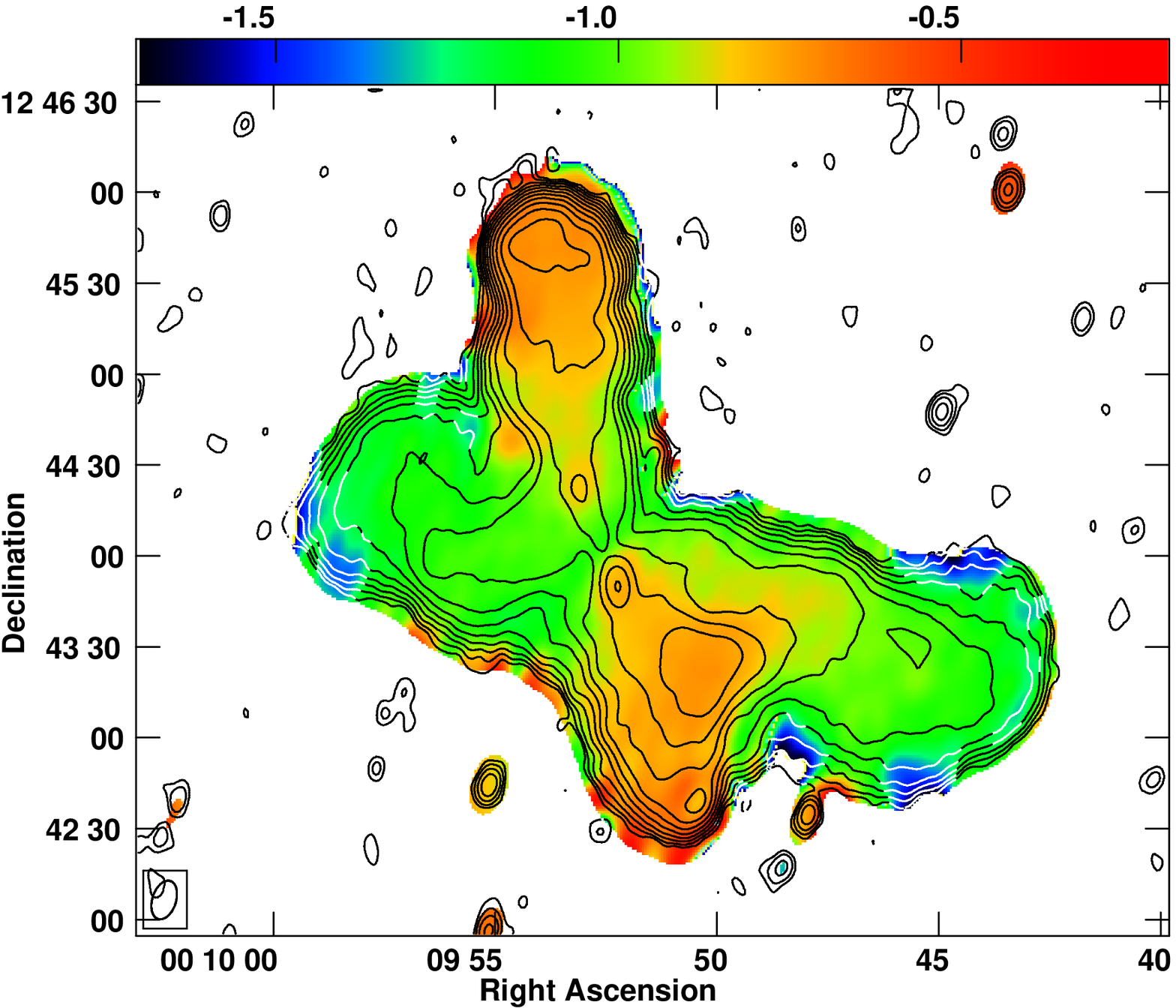} \\
\includegraphics[width=10.8cm]{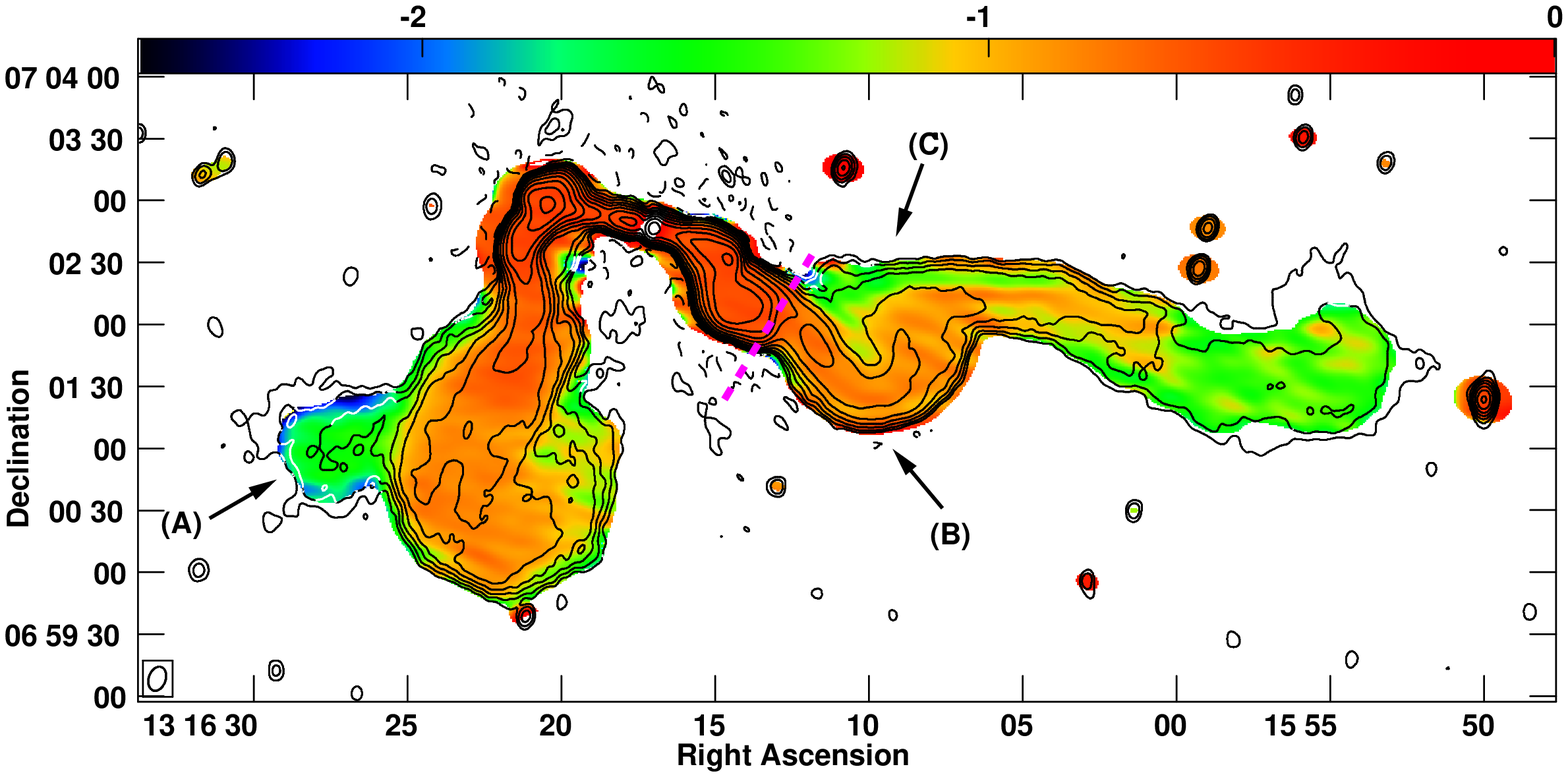} &
\includegraphics[width=6.2cm]{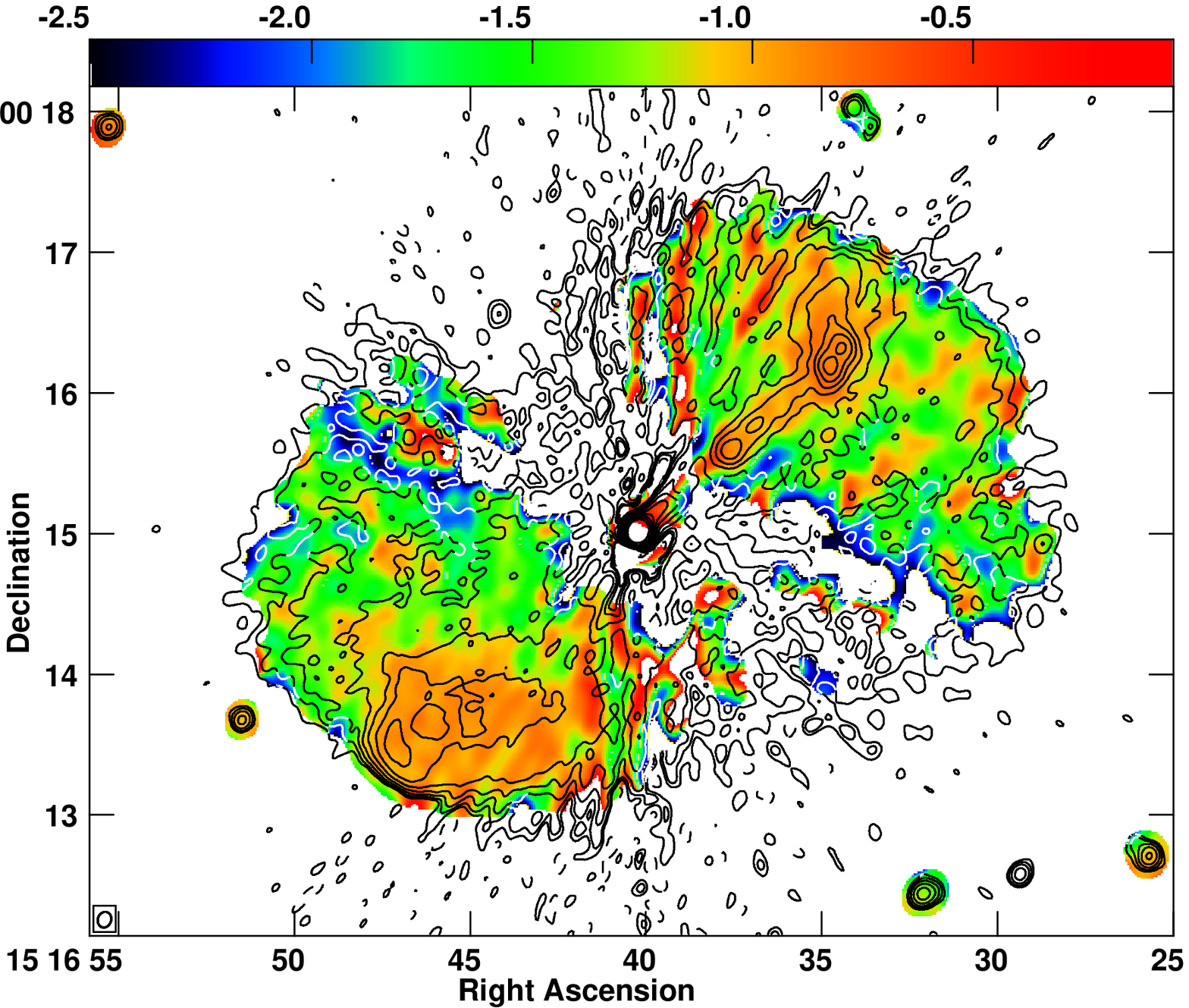}
\end{tabular}
\caption{MeerKAT in-band spectral index images of our four sample sources, 4C\,12.02 (top-left panel), 4C\,12.03 (top-right panel), CGCG\,044$-$046 (bottom-left panel) and CGCG\,021$-$063 (bottom-right panel).
The total intensity radio contours from the MeerKAT data are overlaid on it, with the lowest radio contour plotted is three times the local \textsc{rms} noise and increasing by factors of 2.
We have also marked the three distinct regions of emission beyond the compact flat spectrum core, the inner jets all the way to the hot spots, the central part of the tails, and the terminating part of the tails for the CGCG\,044--046 source.
It highlights three different regions of emission beyond the compact flat spectrum core: (1) the inner jets all the way to the hot spots; (2) the central part of the tails, labelled as `B' and `C', and (3) the terminating part of the tails, labelled as `A'.  This feature, `A' is the end of the tail and is called as the eastern protrusion.
Furthermore the sharp transition from the central to the terminating part of the tail in the western radio emission is marked as dashed magenta line (see also Sec.~\ref{cgcg044}).}
\label{Meer-inband-spix}
\end{center}
\end{figure*}

\subsection{Spectral index imaging}
\label{spec-struc}

We constructed the spectral index image via the standard direct method, {\it i.e.},
$$
log\left(\frac{S_{\nu_1}(x,y)}{S_{\nu_2}(x,y)}\right) \div log\left(\frac{\nu_1}{\nu_2}\right),
$$
where $S_{\nu_1}(x,y)$ and $S_{\nu_2}(x,y)$ are flux densities at pixel location $(x,y)$ for two frequencies, $\nu1$ and $\nu2$.

With this approach we performed MeerKAT in-band spectral index imaging for all sources in our sample.
The MeerKAT datasets were imaged using the WSCLEAN package \citep{wsclean} via a joint deconvolution and 4th order polynomial fitting options of it to make wide-band multi-frequency synthesis images (see also Sec.~\ref{data-reduction}).  This provides images at centre-frequencies and the corresponding in-band spectral index maps. The resulting images are shown in Fig.~\ref{Meer-inband-spix} for all sources, with total intensity radio contours overlaid from the 856--1712\,MHz (L-band) MeerKAT data. 
 
\begin{figure}
\begin{center}
\begin{tabular}{c}
\includegraphics[angle=-90,width=8.05cm]{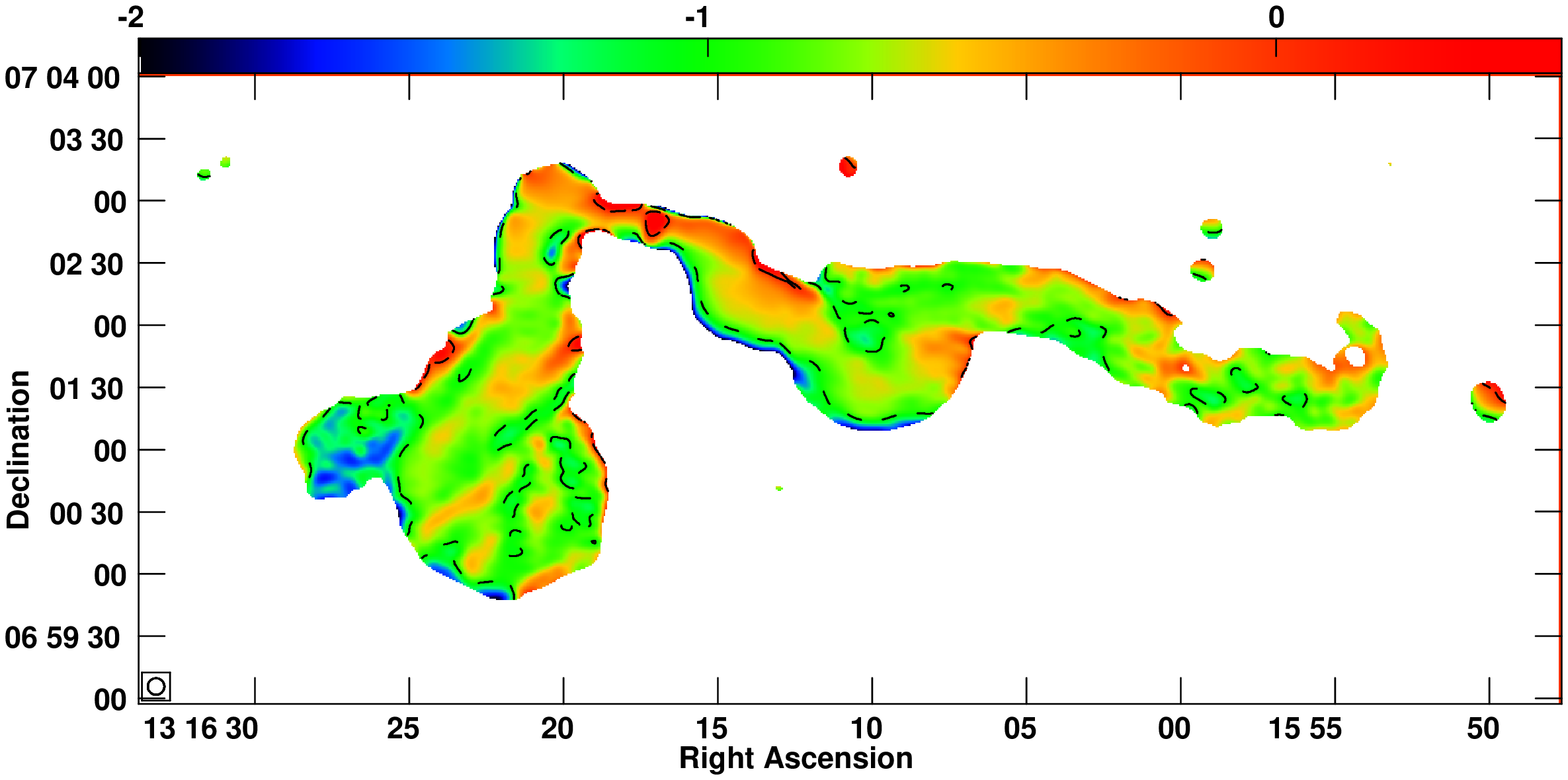}
\end{tabular}
\caption{Spectral index image using the uGMRT and MeerKAT for CGCG\,044$-$046 radio galaxy. The radio contour plotted corresponds to a spectral index of $-$1.0.}
\label{Meer-uGMRT-spix}
\end{center}
\end{figure}

We further produced uGMRT-MeerKAT spectral index images. The uGMRT and MeerKAT full resolution images, which on average give $\theta$ $\approx$ 4\farcs5 and 7\farcs5 at full width half maximum, were restored to the same angular resolution of $\theta$ = 8$^{\prime\prime}$ at both central frequencies for the spectral index imaging.
We point out that the ($u,v$) coverage of the two arrays is nearly identical, hence we did not need to remove baselines in the ($u,v$) planes to match the range of accessible angular scales in the two datasets.
We use these matched resolution images to construct the spectral index images.
The errors in the flux densities are approximately 5\% and 3\% at band-4 and L-band of uGMRT and MeerKAT, respectively, including calibration errors.  The final errors in the spectral indices have been estimated by propagating individual errors in quadrature.
Here we show only the uGMRT-MeerKAT spectral index image which we obtained for CGCG\,044--046 (Figure~\ref{Meer-uGMRT-spix}). For this source the comparable sensitivity of the uGMRT and MeerKAT datasets provides the best result, which adds information to the MeerKAT in-band image. 

CGCG\,021--063 is affected by residual artefacts due to the strong nuclear component, nevertheless the MeerKAT in-band spectral index image is reliable for the strongest features, {\it i.e.} the jets and the hot spots.
The spectral index distribution in our sources (see Fig.~\ref{Meer-inband-spix}) shows a flat spectrum core in all cases.

The spectral index image of 4C\,12.02 is quite puzzling. The spectrum of the very bright hot spots is overall similar to that of the lobes, with $\alpha \simeq -0.8$, suggesting that replenishment of fresh particles is reducing, or has stopped.

Radio galaxies show steepening from the hot spots towards the core, which is consistent with what is found in 4C\,12.03 \citep{Hardcastle2005}, where the lobes are back-flow emission. Our image for 4C\,12.03 shows a clear separation in the spectral index distribution between the north-south and east-west axis, the latter being considerably steeper ($\alpha \simeq -1.2$ to be compared to the values in the range $\alpha \simeq -0.8$ and $-0.9$ in the north-south lobes). Our results are consistent with earlier findings in \citet{Rottmann2001} and \citet{LalandRao2007}.
The source will be further discussed in Sec.~\ref{sec.sum-conc}.

The spectral index image of CGCG\,044--046 shows four clearly separated regions, with the spectral index steepening gradually from the flat to inverted spectrum core, to the ending parts of the tail, where it reaches values of $\alpha \simeq -1.5$.
This source will be further discussed in Sec.~\ref{sec.sum-conc}.

Despite the overall poorer quality of the image, the spectral index distribution of CGCG\,021--063 is quite interesting, with the north-western hot-spot considerably flatter than the surrounding emission from the lobe, {\it i.e.} $\alpha \simeq -0.8$ and $\alpha \simeq -1.2$ respectively. On the other hand, the eastern hot-spot and channels of emission feeding it (see bottom-left panel of Fig.~\ref{Meer-uGMRT}) have a similar spectral index, in the range $\alpha \simeq -0.9$.

\section{Discussion}
\label{sec.sum-conc}

The MeerKAT and uGMRT images presented in this paper cover a frequency range from 550 MHz to 1712 MHz at nearly identical angular resolutions and sensitivities, which is ideal for studying the radio morphology and spectral features in the jets and lobes of radio galaxies. 
The images obtained with the two arrays, for all the sources studied, are remarkably similar in detail, which gives us confidence in the image integrity, imaging processing, and calibration of the uGMRT and MeerKAT data. The remarkable coincidence of the radio contours with the galaxies in the optical overlay (Fig.~\ref{sample-NVSS}) gives us confidence in even the lowest contour in our images.

Overall, our images confirm the morphological classification which is reported in Table~\ref{tab:sample}, and which has been made on the basis of images at lower sensitivity and poorer angular resolution. 
The only exception is CGCG\,021--063, which can be classified as FR\,II radio galaxy.
At the same time, each source has revealed interesting features which open new questions and throw light on the complexity of the interplay of the radio plasma emitted by the host galaxy and the surrounding medium in which the jets propagate. Our total intensity and spectral index imaging suggest that we can broadly identify three different regimes in all our sources: a region close to the AGN (and within the optical host) where the FR\,I and FR\,II division is relevant and fairly clear, a region further from the AGN where the intergalactic medium (IGM) affects the hot spots, and a third region, which goes beyond the hot spots and bears information on the interaction between the tails and the external medium (as in the case of CGCG\,044--046).

Furthermore, the high angular resolution and high sensitivity of our images clearly reveals that substructure in the hot spots is common, which we discuss below.

\subsection{Multiple hot spots}
\label{hotspots}

The western hot spot in 4C\,12.02 has two peaks (labelled in Fig.~\ref{Meer-uGMRT}, upper-right panel), and the one closer to the core is perpendicular to the direction of the lobe.  The eastern hot spot has three peaks along the direction of the jet flow, and it is notable that the brightest peak is not the outermost one.
Some faint emission is also detected beyond the northern hot spot of 4C\,12.03, while the southern hot spot is broad and uniform in surface brightness, which is unusual.
The bent morphology of CGCG\,044--046 is typical of radio galaxies at the centre of groups and clusters of galaxies. At the same time, the structure of the hot spots in the inner part of the source poses the question of what is really bending the jets and how the flow propagates beyond their location.
Finally, the north-western hot spot in CGCG\,021--063 is not the terminating point of the jet, because the radio emission is detected beyond the hot spot. The hot spot itself has two peaks of comparable brightness and is elongated along the direction of propagation of the jet. The south-eastern hot spot is misaligned with the source axis and has complex morphology perpendicular to the jet axis.

Hot spots are believed to be the result of some form of shock, such as the working surface of the jets when they hit the interstellar medium or the IGM, forming the termination shock, also called the standing or reflected shock front \citep{Hardcastle2005}.
The spectral index images shown in Fig.~\ref{Meer-inband-spix} are overall consistent with this idea, but there are some interesting trends that should be noted. On average, the hot spots have flatter spectra than the jets and lobes. However, there are remarkable deviations.
For example, 4C\,12.02 is intriguing, since the spectral index of the hot spots is not different from the emission of the lobes, at least in the MeerKAT frequency range (856--1712 MHz). The integrated spectrum for this source also suggests that the compact features are not dominant in the emission at high frequency.
Additionally, the spectral features of the south-eastern hot spot in CGCG\,021--063 are not different from the spectrum of the emission of the adjacent regions of the lobe.

Radio galaxies with multiple knots and hot spots have been known for quite some time, with the hot spots themselves detected even at wavelengths other than radio wavelengths \citep[e.g.][]{Kraftetal,Hardcastleetal2007,Hardcastleetal2014}.
Our results suggest various possibilities, such as the propagation of the jets through contact discontinuities, or standing shocks.  The sensitivity and resolution of our images allow detailed insight into the brightness distribution of the lobes. Furthermore, each of the sources presented here shows features that deserve attention, which are discussed in the next subsection.

\begin{figure}
\begin{center}
\begin{tabular}{c}
\includegraphics[width=8.05cm]{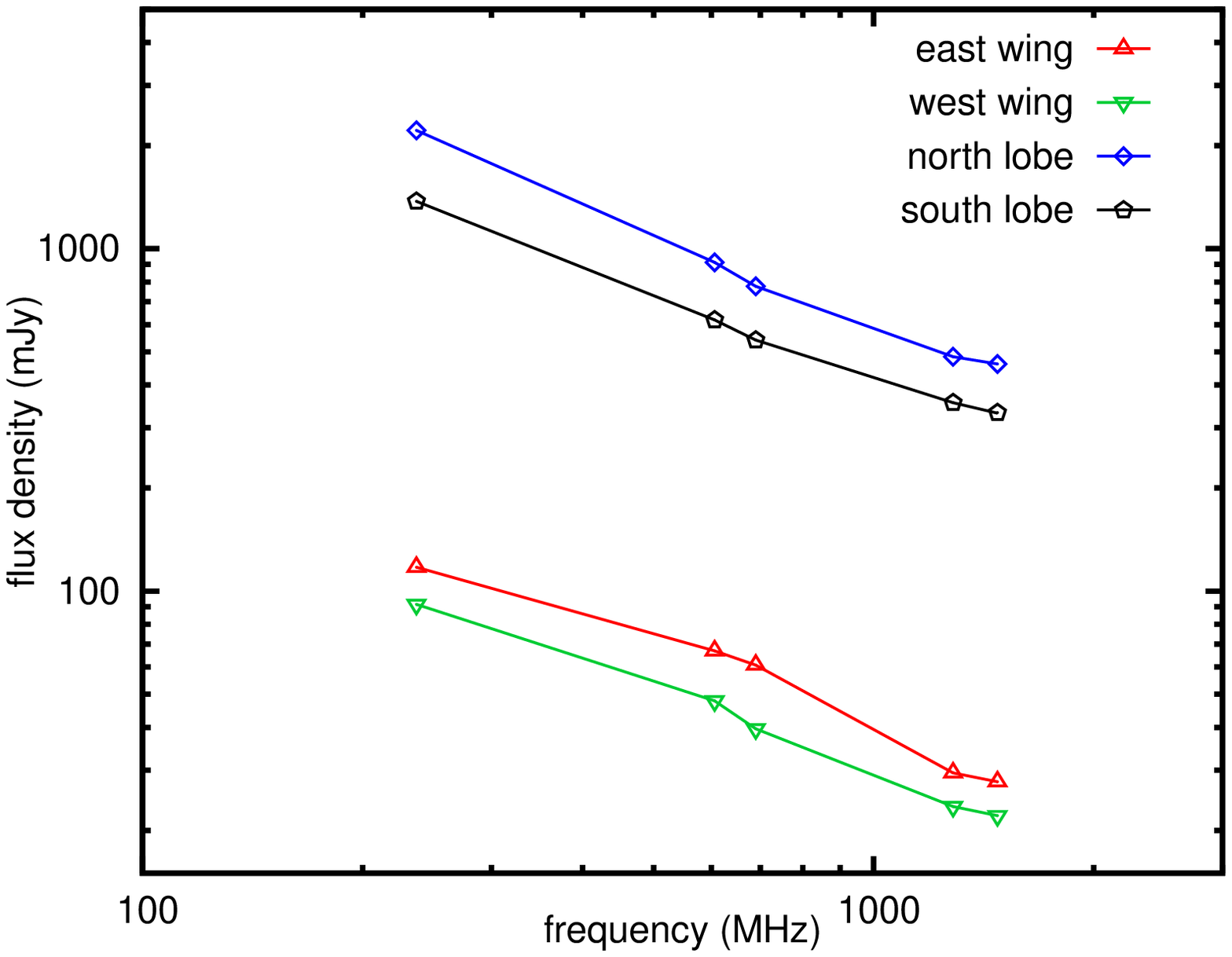}
\end{tabular}
\caption{Flux densities as a function of frequency (spectra) for four distinct regions of 4C\,12.03 radio galaxy at low radio frequencies (see Sec.~\ref{jets}). The 240 MHz and 610 MHz measurements are from \citet{LalandRao2007}, the 690 MHz and 1284 MHz measurements are using uGMRT band-4 and MeerKAT L-band images presented here, and the 1.5 GHz measurements are from \citet{1991AJ....102..537L}.  The error-bars are smaller than the size of the symbols.}
\label{regions-4c12.03}
\end{center}
\end{figure}

\subsection{Jets and lobes}
\label{jets}

The misalignment and asymmetry of the two lobes in 4C\,12.02 is remarkable.  Moreover, the transition between the hot spot and lobe in the western emission is very sharp. We should note that the size of this radio galaxy places it in the class of giant radio galaxies. The properties of the external medium can thus change considerably through almost 800 kpc and projection effects may play an important role \citep{Harwoodetal,Krauseetal}.  Furthermore, despite the morphology with prominent hot spots in both directions of the emission, the uniform spectral index between the lobes and hot spots suggests that the latter are no longer replenished.

The X-shaped morphology of 4C\,12.03, as well as its spectral index distribution, led us to investigate whether the radiative age in the north-south and in the east-west axes differ, as would be expected if they traced two different cycles of radio activity.  The trend of the integrated spectral index in 4C\,12.03, shown in Fig.~\ref{regions-4c12.03}
(green line) highlights the contribution of the east-west structure in the total flux density at high frequencies.
Assuming an equipartition of energy between relativistic particles and magnetic field, we determine magnetic fields, $B_{\rm eq}$ \citep[and minimum energy densities, $u_{\rm eq}$,][]{Miley1980} for the regions encompassing hot spots along the north-south axis and `winged' diffuse radio lobes along the east-west axis. The north and the south hot spots have $B_{\rm eq}$ = 3.8~$\mu$G (and $u_{\rm eq}$ = 13.7 $\times$ 10$^{-13}$ erg~cm$^{-3}$) and $B_{\rm eq}$ = 3.1 $\mu$G (and $u_{\rm eq}$ = 9.5 $\times$ 10$^{-13}$ erg~cm$^{-3}$) respectively, whereas the east and the west diffuse lobes have $B_{\rm eq}$ = 2.1 $\mu$G (and $u_{\rm eq}$ = 4.5 $\times$ 10$^{-13}$ erg~cm$^{-3}$) and $B_{\rm eq}$ = 1.4 $\mu$G (and $u_{\rm eq}$ = 2.0 $\times$ 10$^{-13}$ erg~cm$^{-3}$), respectively.
Here we have assumed the ratio of energy in the heavy particles to that in the electrons and the filling factor of the emitting regions to be unity.  The sizes of the four regions are symmetrical rectangular regions, each being at least $\sim$15--20 beams across.
From Fig.~\ref{regions-4c12.03} we estimated a break frequency $\nu_{\rm br}$=0.69 GHz for the east and west wings, and assumed $\nu_{\rm br}$=1.2 GHz for the north and south lobes.
This leads to radiative lifetimes \citep{JaffePerola1973} of 40.8 Myr and 41.4 Myr respectively for the north and the south lobes (this value should be considered an upper limit, as the assumed break frequency is a lower limit),  and 52.1 Myr and 51.5 Myr for the east and west diffuse lobes, respectively.
The estimates of both the equipartition magnetic field
and the radiative age, though not conclusive, are consistent with the possibility that the east and west diffuse wings are formed due to hydrodynamic mechanism, {\it i.e.} back-flow from the north and south active lobes, as suggested also by the in-band spectral index map of 4C\,12.03 (Fig.~\ref{Meer-inband-spix}, top-right panel).

4C\,12.03 is also characterised by two peaks of emission along the north-south jet, symmetrically located with respect to the core and separated by $\sim$ 80 kpc (Figs.~\ref{sample-NVSS} and \ref{Meer-uGMRT}). They are well visible also in the spectral index distribution (Fig.~\ref{Meer-inband-spix}).
They could be either brightness peaks of the underlying jets feeding the hot spots, or an inner double source related to a new cycle of radio activity \citep{Schoenmakersetal}. Unfortunately, our data do not allow us to discriminate between the two alternatives.
Finally, we point out that the active axis is not aligned with the axis of the inner double source. The northern hot spot is aligned at $\sim$10 deg, and the southern one at $\sim$40 deg. It is unclear what could be responsible for such misalignment over scales of several hundreds of kiloparsec.

\begin{figure*}
\begin{center}
\begin{tabular}{c}
\includegraphics[angle=-90,width=15.8cm]{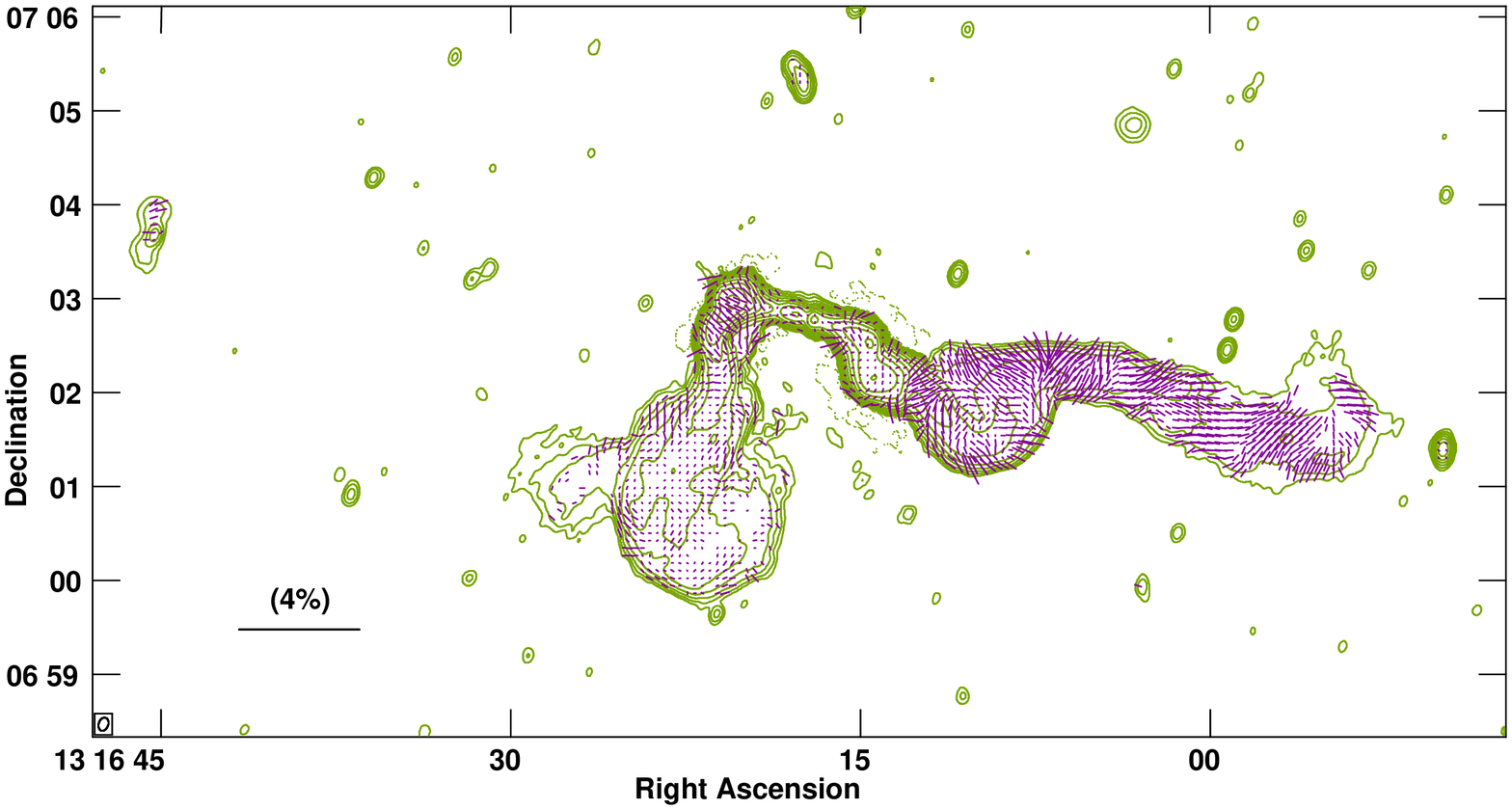} \\
\includegraphics[angle=-90,width=15.8cm]{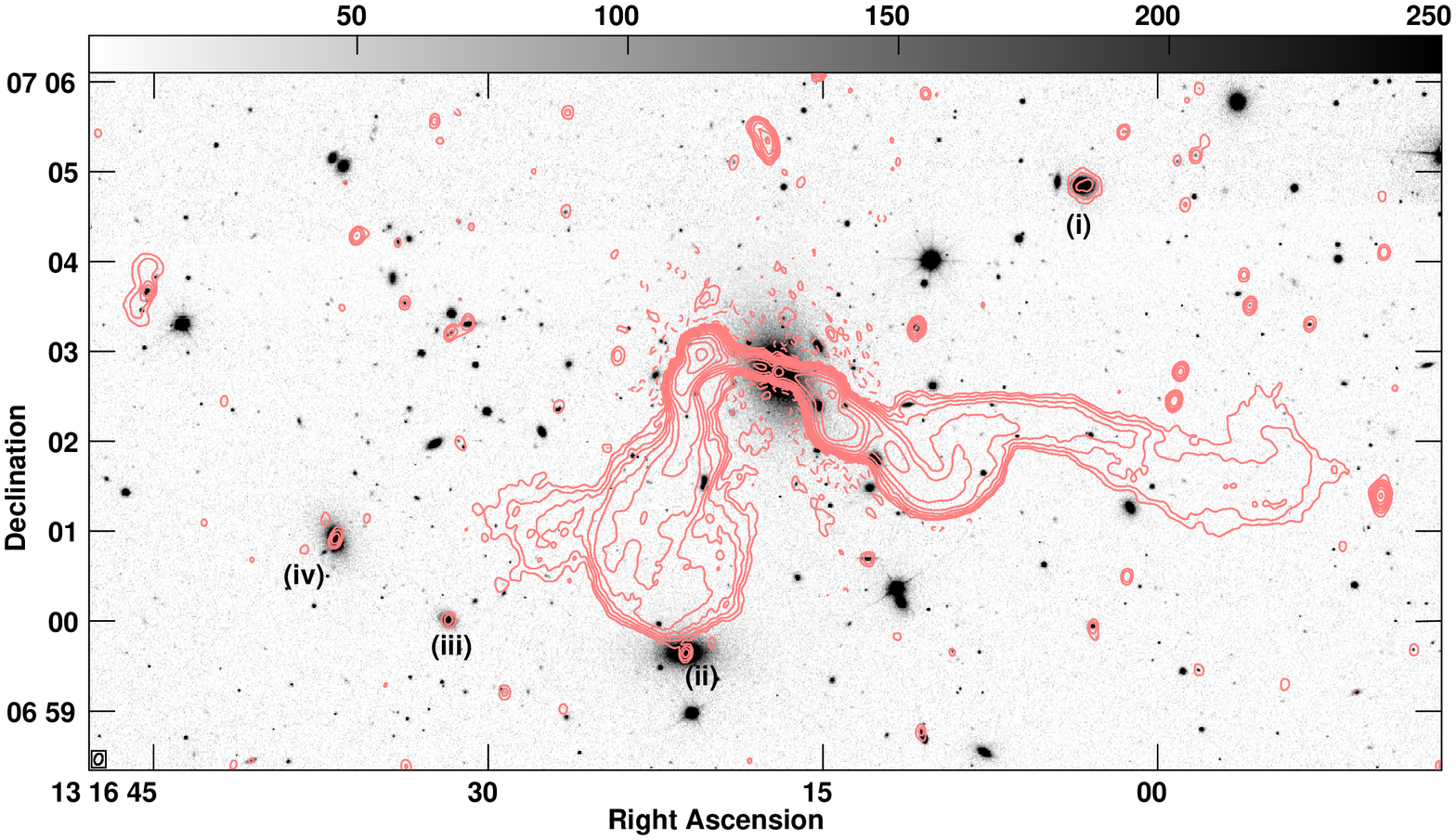}
\end{tabular}
\caption{The radio polarisation image (upper-panel) and the $r$-band SDSS (DR12) image (lower-panel) of CGCG\,044$-$046 with the MeerKAT L-band total intensity radio contours overlaid on them.
Also overlaid on the polarisation image are the EVPA orientation vectors.  In both images, the lowest radio contour plotted is approximately three times the local \textsc{rms} noise (= 18.0~$\mu$Jy~beam$^{-1}$ and 23.5~$\mu$Jy~beam$^{-1}$, in upper and lower panels, respectively) and increasing by factors of 2.
The distribution of \textbf{E} vector position angles are shown, where the vectors have lengths proportional to the degree of polarisation $p$.  The length of the vector, 1~arcsec corresponds to 3.125\% (the length of the vector corresponding to p = 4\% is shown in the lower left corner).  No vectors are shown where the polarised signal-to-noise is $<$4:1. The vectors are shown slightly undersampled, at $\sim$5~arcsec intervals, for clarity.  Four bright galaxies that are also part of the Zwicky cluster 1313.7$+$0721, of which CGCG\,044$-$046 is the dominant galaxy, are also marked (see also Sec.~\ref{cgcg044}).}
\label{cgcg-polar}
\end{center}
\end{figure*}

Unfortunately, the artefacts affecting the images of CGCG\,021--063 do not allow a detailed analysis of its features.  However, the filamentary structure of the south-eastern lobe is clear and remarkable. The uGMRT total intensity image is suggestive of two parallel channels feeding the hot spots.  Alternatively, they could be the outer walls of the propagating continuous stream of radio-emitting synchrotron plasma (labelled in Fig.~\ref{Meer-uGMRT}, upper-left panel).
Looking more closely at the uGMRT and MeerKAT images together, the
two parallel channels in the south lobe are off the line joining the north hot spot through the radio core to the south hot spot, which contradicts the simple picture that the two parallel channels in the south lobe could be feeding the hot spots.  It seems more likely that the two channels are leading away from the south hot spot, rather than leading towards it, which then suggests that these parallel channels are probably the reflected shocks or back-flow emission.
The MeerKAT in-band spectral index in this region is very uniform, $\simeq$ $-$0.9, with no apparent transition between the hot spot and the emission feeding it.
The jets and hot spots are surrounded by a cocoon and the cocoon could be a result of back-flow emission.
It is possible that projection effects play a role in this source, as suggested by the very compact core and the strong asymmetry between the north-east and south-west jets.

\subsection{CGCG 044--046}
\label{cgcg044}

The overall observational properties of CGCG\,044--046 source suggest that the local environment plays a major role.  More specifically, the physical mechanism giving rise to bent tails is possibly due to the motion of the host galaxy through the ICM.  The source extends $\sim$ 414 kpc to the west and $\sim$ 207 kpc to the south-east, {\it i.e.} well into the ICM.
The source is remarkably similar to 3C\,465, the prototype of wide-angle tail radio galaxy \citep{Eileketal,Giacintuccietal2007}.

The total intensity and spectral index images highlight three different regions of emission beyond the compact flat spectrum core: (1) the inner jets all the way to the hot spots; (2) the central part of the tails and (3) the terminating part of the tails.
The overlay of the MeerKAT contours and the $r$-band SDSS image of the optical field shows that the hot spots are located outside the envelope of the optical galaxy, at least at the sensitivity level of the $r$-band SDSS image. We point out that the m$_g$ = 16.7 galaxy just north of CGCG\,044--046 and within its optical envelope (at least in projection) is VIII Zw 276 and is located at the cluster redshift, $z$ = 0.049567.

The inner jets and the hot spots (region 1) have a fairly uniform spectral index, $\simeq$ $-$0.6, which is quite flat. The transition between the inner jet and the central part of the tail is sharp in the western part of the radio galaxy, both in the total intensity image and in the spectral index, which steepens to $\alpha$ $\simeq$ $-$0.9.  This part of the tail (region 2) shows prominent bending with edge brightening.  A sharp spectral transition with similar values is seen beyond the eastern hot spot, too. In the eastern lobe, we note a flatter central ridge with $\alpha\approx-0.7$ surrounded by steeper emission with $\alpha\approx-0.9$.

In the western emission, the transition from the central to the terminating part of the tail (region 3) is sharp both in the total intensity image and spectral index, which  steepens from values around $-$1.0 to values around $-$1.5. The eastern lobe, on the other hand, seems to bend behind the emission itself, and we believe that the eastern protrusion (labelled `A' in Fig.~\ref{Meer-inband-spix}) with a spectral index as steep as $-$1.6/$-$1.8 is the end of the tail.

The origin of these sharp transitions in surface brightnesses and spectral properties is unclear.
They could reflect intermittent activity in the radio emission of the AGN and/or significant interaction of the jets and lobes with discontinuities in the IGM.

Further hints of substantial bending come from the polarisation properties. This source is the only one with polarisation calibration source included in our MeerKAT observation.
\citet{Patnaik1986} presented a polarisation study at higher frequency (VLA $L$- and $C$-band), and was able to constrain rotation measures across the source to within $\pm30\,\mathrm{rad}\,\mathrm{m}^{-2}$.  Although these values are under the detectable range with our frequency coverage, we used full-band $Q$ and $U$ maps and did not perform rotation measure analysis.

The upper panel of Fig.~\ref{cgcg-polar} shows the electric field vectors superimposed on the MeerKAT total intensity image. There is a remarkable asymmetry in polarisation between the western and eastern regions of the radio emission and the two tails differ significantly in polarisation fraction. In particular, the central and terminating part of the western tail is much more polarised than the south-eastern one. Our recovered polarisation vectors are consistent with \citet{Patnaik1986}, but provide more detail across the lobes. Considering that this radio galaxy is at the centre of a galaxy cluster, where dense ICM is usually found, we suggest that the source is not in the plane of the sky and interpret the polarisation asymmetry as due to Laing-Garrington effect \citep{Garringtonetal}, {\it i.e.}, an external medium that is denser against the south-eastern lobe.

As a final remark, we point out that the four bright galaxies with radio emission clearly visible in the lower panel of Fig.~\ref{cgcg-polar} are all part of the same Zwicky cluster 1313.7$+$0721. In particular, using data gleaned from NED (labelled in Fig.~\ref{cgcg-polar}, lower panel), their names from west to east are as follows:
(i) WISEA\,J131604.49$+$070453.4 (m$_g$=18.4) at $z$ = 0.049298;
(ii) CGCG\,044$-$047 (m = 15.6) at $z$ = 0.049551;
(iii) WISEA\,J131631.79$+$070000.9 (no magnitude available) at $z$ = 0.049297; and
(iv) WISEA\,J131636.85$+$070054.6 (m = 17.11) at $z$ = 0.047596.

\section{Concluding remarks}
\label{conc-rem}

In this paper we have presented uGMRT and MeerKAT images of four radio galaxies belonging to a larger sample of 12 FR\,I and FR\,II radio galaxies selected from the 4C catalogue with the main goal of investigating whether the sharp differences between the FR\,I and FR\,II morphologies still hold with the improved imaging capabilities of the current generation of radio interferometers, or if we need more morphological classes.

We have explored the radio morphology of our four targets - 4C\,12.02, 4C\,12.03, CGCG\,044--046 and CGCG\,021--063 - in the light of the superb image sensitivity reached by the uGMRT in the 550--850 MHz band, and by the MeerKAT in the 856--1712 MHz band. The collected information hence seamlessly spans more than 1 GHz in bandwidth.
We have supplemented our total intensity datasets with MeerKAT in-band spectral imaging for all sources. Furthermore, for CGCG\,044--046 we obtained the uGMRT-MeerKAT spectral structure and MeerKAT polarisation information.

The MeerKAT and uGMRT images of all four radio galaxies presented here are remarkably similar in detail, which gives us confidence in image fidelity, imaging processes and calibration of the uGMRT and MeerKAT. Moreover, in all cases, the morphology in the uGMRT band-4 and in the MeerKAT L-band is the same, even in the finest details.

While we conclude that the overall FR\,I--FR\,II classification scheme still holds, at least for our targets, the combination of $\mu$Jy~beam$^{-1}$ sensitivity and high ($\sim 5^{\prime\prime}$ to 7$^{\prime\prime}$) angular resolution over the full 550-1712 MHz range reveals very interesting features.
For example, filamentary emission in the lobes and substructure in the hot spots is common.  Moreover, in CGCG\,044--046, CGCG\,021--063 and in the north hot-spot of 4C\,12.03 the radio emission extends beyond the hot spots themselves.
In 4C\,12.02 and 4C\,12.03 the two hot-spots are not at 180$^{\circ}$ with respect to the core, and it is unclear what may cause such slight misalignment over hundreds of kpc.

The MeerKAT in-band spectral imaging has complemented and confirmed the morphological classification from the total intensity radio images and provided new insights into some of the radio sources. In particular, the uniform steep spectrum in the lobes and hot-spots of 4C\,12.02 suggests that the hot-spots are no longer replenished by fresh electrons. The radio emission in the east-west axis of 4C\,12.03 is steeper than the north-south axis, and one possibility is that the former is the result of hydrodynamic back-flow, as has been recently seen in PKS\,2014--55 \citep{Cottonetal}. This radio galaxy also shows two inner brightness peaks, which form an inner double $\simeq$80 kpc in size, aligned with the north-south axis of emission. Further investigation will be required to determine if this is the signature of restarted activity.

Finally, the MeerKAT in-band spectral structure and polarisation information for CGCG\,044--046 shows that most likely the source is not in the plane of the sky and that the observed properties are strongly affected by the cluster environment. We interpret the asymmetries in the polarisation properties as due to the Laing--Garrington effect.

Our work further shows that very good image sensitivity over a broad range of angular scales is necessary to perform a detailed study of radio galaxies, and warns against the accuracy of the morphological classification of radio galaxies made with source detection algorithms, especially for surveys made with radio telescopes which have much lower sensitivity to diffuse, low brightness emission.

At least for the sources presented in this paper, we conclude that there is probably little value in establishing a more complicated morphological classification system.  It appears to us that the FR\,I and FR\,II (and possibly FR\,0) morphologies are common and basic and that the complexities are introduced by the local environment, buoyancy and motions through the IGM. All are probably at play, together with reflection, refraction or bifurcation of bow shocks at discontinuities in the IGM. Adding further descriptors (e.g. wide-angle tail or narrow-angle tail) may be useful to link the morphology to the environment in which the lobes are formed or propagate, whereas the FR\,I/II classification probably describes conditions closer to or more intrinsic to the AGN and host galaxy.

\section*{Acknowledgments}

We thank the anonymous referee for his/her comments that improved and broadened the scope of this paper.
DVL acknowledges the support of the Department of Atomic Energy, Government of India, under project no. 12-R\&D-TFR-5.02-0700.
TV acknowledges the support from the Ministero degli Affari Esteri e della  Cooperazione Internazionale, Direzione Generale per la Promozione del Sistema Paese, Progetto di Grande Rilevanza ZA18GR02.
OS's research is supported by the South African Research Chairs Initiative of the Department of Science and Technology and National Research Foundation.
DK acknowledges funding from the European Research Council (ERC) under the European Union’s Horizon 2020 research and innovation programme (grant agreement no. 679627).
FL acknowledges financial support from the Italian Minister for Research and Education (MIUR), project FARE, project code R16PR59747, project name FORNAX-B.
SVW acknowledges the financial assistance of the South African Radio Astronomy Observatory.
We thank the staff of the GMRT who made these observations possible. 
The GMRT is run by the National Centre for Radio Astrophysics of the Tata Institute of Fundamental Research.
The MeerKAT telescope is operated by the South African Radio Astronomy Observatory, which is a facility of the National Research Foundation, an agency of the Department of Science and Innovation.
We are grateful to the full MeerKAT team at SARAO for their work on building and commissioning MeerKAT.
This research has made use of the NED, which is operated by the Jet Propulsion Laboratory, Caltech, under contract with the NASA, and NASA's Astrophysics Data System.

\section*{Data Availability}

The GMRT data underlying this article are available via the GMRT online archive
facility\footnote{\url{https://naps.ncra.tifr.res.in/goa/data/search
[naps.ncra.tifr.res.in]}}. The MeerKAT data will be publicly available via the SARAO
archive\footnote{\url{https://archive.sarao.ac.za [archive.sarao.ac.za]}} (proposal ID
SCI-20190418-BF-01) after the end of the proprietary period in mid-2021, but may be shared
earlier upon reasonable request to the corresponding author. All data analyses packages used
in this work are publicly available, and their URLs have been noted in the main text.

\bsp    
\label{lastpage}

\begin{thebibliography}{}
\expandafter\ifx\csname natexlab\endcsname\relax\def\natexlab#1{#1}\fi

\bibitem[Baldi, Capetti \& Giovannini(2019)]{Baldietal} Baldi, R. D., Capetti, A., Giovannini, G. 2019, MNRAS, 482, 2294

\bibitem[Bassani et~al.(2016)]{2016MNRAS.461.3165B} Bassani, L., Venturi, T., Molina, M., et~al. 2016, MNRAS, 461, 3165

\bibitem[Bautz \& Morgan(1970)]{BM70} Bautz, L. P., Morgan, W. W. 1970, ApJ 162, L149

\bibitem[Best \& Heckman(2012)]{BestHeckman} Best, P. N., Heckman, T. M. 2012 MNRAS 421, 1569

\bibitem[Bicknell et~al.(2004)]{Bicknelletal} Bicknell, G., Jones, D. L., Lister, M., et al. 2004 NewAR 48, 1151


\bibitem[Brunetti \& Jones(2014)]{BrunettiJones2014} Brunetti, G., Jones, T. W. 2014, Int. J. Mod. Phys. D, 23, 30007

\bibitem[Bruni et~al.(2019)]{2019ApJ...875...88B} Bruni, G., Panessa, F., Bassani, L., et al. 2019, ApJ, 875, 88

\bibitem[Burns et~al.(1987)]{Burnsetal} Burns, J. O., Gisler, G. R., Borovsky, J. E., et~al. 1987 AJ 94, 587

\bibitem[Capetti et~al.(2020)]{2020A&A...642A.107C} Capetti, A., Brienza, M., Baldi, R. D., et~al. 2020, A\&A, 642, A107 

\bibitem[Condon et~al.(1998)]{Condonetal} Condon, J. J., Cotton, W. D., Greisen, E. W., et~al. 1998 AJ 115, 1693

\bibitem[Cotton et al.(2020)]{Cottonetal} Cotton, W. D, Thorat, K., Condon, J. J., et al. 2020, MNRAS 495, 1271

\bibitem[Croston et~al.(2019)]{Croston2019} Croston, J. H., Hardcastle, M. J., Mingo, B., et~al. 2019, A\&A 622A, 10

\bibitem[Croston et~al.(2017)]{Croston2017} Croston, J. H., Ineson, J., Hardcastle, M. J., Mingo, B. 2017, MNRAS 470, 1943

\bibitem[Dabhade et~al.(2020)]{2020A&A...635A...5D} Dabhade, P., R\"ottgering, H. J. A., Bagchi, J., et~al. 2020, A\&A, 635, A5

\bibitem[Dennett-Thorpe et~al.(2002)]{Dennett2002} Dennett-Thorpe, J., Scheuer, P. A. G., Laing, R. A., et~al. 2002, MNRAS, 330, 60

\bibitem[Dodson et al.(2008)]{Dodsonetal} Dodson, R., Fomalont, E. B., Wiik, K., et al. 2008, ApJS, 175, 314

\bibitem[Eilek et al.(1984)]{Eileketal} Eilek, J. A., Burns, J. O., O'Dea, C. P., et al. 1984, ApJ, 278, 37

\bibitem[Fanaroff \& Riley(1974)]{FanaroffRiley} Fanaroff, B. L., Riley, J. M. 1974, MNRAS, 167, 31P


\bibitem[Garon et~al.(2019)]{Garonetal} Garon, A. F., Rudnick, L., Wong, O. I., et al. 2019 AJ, 157, 126

\bibitem[Garrington et~al.(1988)]{Garringtonetal} Garrington, S. T., Leahy, J. P., Conway, R. G., et al. 1988, Nature, 331, 147

\bibitem[Giacintucci et~al.(2007)]{Giacintuccietal2007} Giacintucci, S., Venturi, T., Murgia, M., et~al. 2007, A\&A, 476, 99



\bibitem[Giovannini et al.(2001)]{Giovanninietal} Giovannini, G., Cotton, W. D., Feretti, L., et al. 2001 ApJ 552, 508

\bibitem[Gong, Li \& Zhang(2011)]{2011ApJ...734L..32G} Gong, B. P., Li, Y. P., Zhang, H. C. 2011, ApJL, 734, L32

\bibitem[Gower, Scott \& Wills(1996)]{1967MmRAS..71...49G} Gower, J. F. R., Scott, P. F., Wills, D. 1967, MmRAS, 71, 49

\bibitem[Gregory \& Condon(1991)]{1991ApJS...75.1011G} Gregory, P. C., Condon, J. J. 1991, ApJS, 75, 1011

\bibitem[Griffith et~al.(1995)]{1995ApJS...97..347G} Griffith, M. R., Wright, A. E., Burke, B. F., Ekers, R. D. 1995, ApJS, 97, 347

\bibitem[Gupta et~al.(2017)]{Guptaetal2017} Gupta, Y., Ajithkumar, B., Kale, H. S., et~al. 2017, Current Science, 113, 707


\bibitem[Hardcastle \& Croston(2020)]{HardcastleCroston} Hardcastle, M. J., Croston, J. H. 2020, NewAR, 88, 1539

\bibitem[Hardcastle et~al.(2019)]{Hardcastlen326} Hardcastle, M. J., Croston, J. H., Shimwell, T. W., et~al. 2019, MNRAS, 488, 3416

\bibitem[Hardcastle \& Krause(2014)]{Hardcastleetal2014} Hardcastle, M. J., Krause, M. G. H. 2014, MNRAS, 443, 1482

\bibitem[Hardcastle et~al.(2007)]{Hardcastleetal2007} Hardcastle, M. J., Croston, J. H., Kraft, R. P. 2007, ApJ, 669, 893

\bibitem[Hardcastle(2005)]{Hardcastle2005} Hardcastle, M. J. 2005, Phil. Trans. R. Soc., 363, 2711

\bibitem[Harwood, Vernstrom \& Stroe(2020)]{Harwoodetal} Harwood, J. J., Vernstrom, T., Stroe, A. 2020, MNRAS, 491, 803 

\bibitem[Heckman \& Best(2014)]{HeckmanBest} Heckman, T. M., Best, P. N. 2014, ARA\&A 52, 589

\bibitem[Heckman et~al.(1994)]{Heckmanetal} Heckman, T. M., O'Dea, C. P., Baum, S. A., Laurikainen, E. 1994, ApJ, 428, 65

\bibitem[Hugo et~al.(2021)]{tricolour}{Hugo, B. V., et~al. 2021, in Astronomical Data Analysis Software \& Systems XXX, ASP Conf. Ser., in press}

\bibitem[Hurley-Walker et al.(2017)]{gleametal}
Hurley-Walker, N., Callingham, J. R., Hancock, P. J., et al., 2017, MNRAS, 464, 1146

\bibitem[Intema et al.(2017)]{Intemaetal} Intema, H. T., Jagannathan, P., Mooley, K. P., Frail, D. A. 2017, A\&A, 598, A78

\bibitem[Ishwara-Chandra \& Saikia(1999)]{1999MNRAS.309..100I}  Ishwara-Chandra, C. H., Saikia, D. J. 1999, MNRAS, 309, 100

\bibitem[Jaffe \& Perola(1973)]{JaffePerola1973} Jaffe, W. J., Perola, G. C. 1973, A\&A, 26, 423


\bibitem[Jonas \& MeerKAT Team(2016)]{JonasandMeerKAT} Jonas \& the MeerKAT team 2016, MeerKAT Science: On the Pathway to the SKA, 1

\bibitem[J\'ozsa et~al.(2021)]{caracal2}{J\'ozsa, G. I. G., White, S. V., Thorat, K., et al. 2021, in Astronomical Data Analysis Software \& Systems XXX, ASP Conf. Ser., in press}

\bibitem[J\'ozsa et~al.(2020)]{caracal1}{J\'ozsa, G. I. G., White, S. V., Thorat, K., et~al. 2020, in Astronomical Data Analysis Software \& Systems XXIX, ASP Conf. Ser., 527, 635}

\bibitem[Kenyon et~al.(2018)]{cubical} Kenyon J. S., Smirnov O. M., Grobler T. L., Perkins S. J. 2018, MNRAS, 478, 2399

\bibitem[Kraft et~al.(2005)]{Kraftetal} Kraft, R. P., Hardcastle, M. J., Worrall, D. M., Murray, S. S. 2005, ApJ, 622, 149

\bibitem[Krause, Hardcastle \& Shabala(2019)]{Krauseetal} Krause, M. G. H., Hardcastle, M. J., Shabala, S. S., 2019, A\&A, 627A, 113

\bibitem[Laing \& Bridle(2012)]{LaingBridle2012} Laing, R. A., Bridle, A. H. 2012 MNRAS 424, 1149



\bibitem[Laing, Riley \& Longair(1983)]{1983MNRAS.204..151L} Laing, R. A., Riley, J. M., Longair, M. S. 1983, MNRAS, 204, 151

\bibitem[Lal(2020)]{Lal2020} Lal, D. V. 2020, ApJS, 250, 22

\bibitem[Lal \& Rao(2007)]{LalandRao2007} Lal, D. V.,  Rao, A. P. 2007, MNRAS, 374, 1085


\bibitem[Large et~al.(1981)]{1981MNRAS.194..693L} Large, M. I., Mills, B. Y., Little, A. G., Crawford, D. F., Sutton, J. M. 1981, MNRAS, 194, 693

\bibitem[Leahy \& Perley(1991)]{1991AJ....102..537L} Leahy, J. P., Perley, R. A. 1991, AJ, 102, 537

\bibitem[Lehmensiek \& Theron(2014)]{Lehmensiek2014} Lehmensiek, R. \& Theron, I. P. 2014, in 31st URSI General Assembly and Scientific Symposium (GASS), pp. 1–4

\bibitem[Lehmensiek \& Theron(2012)]{Lehmensiek2012} Lehmensiek, R. \& Theron, I. P. 2012, in Proc. Int. Conf. Electromagn. Adv. Appl. (ICEAA), pp. 321–324

\bibitem[Maccagni et~al.(2020)]{Maccagnietal} Maccagni, F. M., Serra, P., Murgia, M., et~al. Proc. IAU Symposium No. 359, Galaxy Evolution and Feedback Across Different Environments, eds.

\bibitem[Mahatma et~al.(2019)]{2019A&A...622A..13M} Mahatma, V. H., Hardcastle, M. J., Williams, W. L., et~al. 2019, A\&A, 622, A13

\bibitem[Merritt \& Ekers(2002)]{MerrittEkers} Merritt, D., Ekers, R. D. 2002, Science, 297, 131

\bibitem[Miley(1980)]{Miley1980} Miley, G. K. 1980, ARA\&A, 18, 165

\bibitem[Mingo et~al.(2019)]{Mingoetal} Mingo, B., Croston, J. H., Hardcastle, M. J., et al. 2019, MNRAS, 488, 2701


\bibitem[Nolting et~al.(2019)]{Noltingetal} Nolting, C., Jones, T. W., O'Neill, J. B., et al. 2019, ApJ, 876, 154 

\bibitem[Noordam \& Smirnov(2010)]{meqtrees} Noordam, J. E., Smirnov, O. M., 2010, A\&A, 524, A61

\bibitem[O'Dea \& Owen(1985)]{ODeaOwen1985} O'Dea, C. P., Owen, F. N. 1985, AJ, 90, 927

\bibitem[O'Donoghue, Owen \& Eilek(1990)]{ODonoghueetal1990} O'Donoghue, A, Owen, F. N., Eilek, J. A. 1990, ApJS, 72, 75


\bibitem[Offringa et~al.(2014)]{wsclean} Offringa, A. R., McKinley, B., Hurley-Walker, N., et~al. 2014 MNRAS, 444, 606



\bibitem[Parkes Catalog (1990)]{1990PKS90.C...0000W} Parkes Catalog, 1990, Australia telescope national facility

\bibitem[Patnaik, Malkan \& Salter(1986)]{Patnaik1986} Patnaik, A. R., Malkan, M. A., Salter, C. J. 1986, MNRAS, 220, 351

\bibitem[Patnaik, Banhatti \& Subrahmanya(1984)]{Patnaik1984} Patnaik, A. R., Banhatti, D. G., Subrahmanya, C. R. 1984, MNRAS, 211, 775

\bibitem[Perley \& Butler(2013)]{PerleyButler} Perley, R. A., Butler, B. J. 2013, ApJS 204, 19

\bibitem[Pilkington \& Scott(1965)]{PilkingtonScott} Pilkington, J. D. H., Scott, P. F. 1965, MmRAS, 69, 183

\bibitem[Proctor(2016)]{Proctor} Proctor, D. D. 2016, ApJS, 224, 18

\bibitem[Ramatsoku et~al.(2020)]{Ramatsokuetal} Ramatsoku, M., Murgia, M., Vacca, V., et~al. 2020, A\&A, 636, L1


\bibitem[Richards et~al.(2014)]{2014MNRAS.438.3058R} Richards, J. L., Hovatta, T., Max-Moerbeck, W., et~al. 2014, MNRAS, 438, 3058

\bibitem[Rottmann(2001)]{Rottmann2001} Rottmann, H., 2001, PhD thesis, Rheinischen Friedrich-Wilhelms-Universit\"at Bonn

\bibitem[Rudnick \& Owen(1976)]{RudnickOwen1976} Rudnick, L, Owen, F. N. 1976, ApJ, 203, L107


\bibitem[Schellenberger et~al.(2017)]{Schnellenbergetal} Schellenberger, G., Vrtilek, J. M., David, L., et~al. 2017, ApJ, 845, 84

\bibitem[Schilizzi \&  McAdam(1975)]{1975MmRAS..79....1S} Schilizzi, R. T.,  McAdam, W. B. 1975, MmRAS, 79, 1

\bibitem[Schoenmakers et~al.(2000)]{Schoenmakersetal} Schoenmakers, A. P.,  Mack, K. -H.,  de~Bruyn, A. G.,  et~al. 2000, A\&A, 146, 293

\bibitem[Slee(1995)]{1995AuJPh..48..143S} Slee, O. B. 1995, AuJPh, 48, 143

\bibitem[Swarup et al.(1991)]{Swarupetal1991} Swarup, G., Ananthakrishnan, S., Kapahi, V. K., et al. 1991, Current Science, 60, 95


\bibitem[Venturi et~al.(1995)]{Venturietal1995} Venturi, T., Castaldini, C., Cotton, W. D., et~al. 1995 ApJ 454, 735



\bibitem[White et~al.(2020a)]{2020PASA...37...18W} White, S. V., Franzen, T. M. O.,  Riseley, C. J., et al. 2020a, PASA, 37, 18

\bibitem[White et~al.(2020b)]{2020PASA...37...17W} White, S. V., Franzen, T. M. O.,  Riseley, C. J., et al. 2020b, PASA, 37, 17

\bibitem[White \& Becker(1992)]{1992ApJS...79..331W} White, R. L., Becker, R. H. 1992, ApJS, 79, 331

\bibitem[White, Becker \& Edwards(1991)]{1991ApJS...75....1B} White, R. L., Becker, R. H., Edwards, A. L. 1991, ApJS, 75, 1

\bibitem[Wright \& Otrupcek(1981)]{1981A&AS...45..367K} Wright, A., Otrupcek, R. Parkes Catalog, 1990, Australia telescope national facility

\bibitem[Yang et al. (2019)]{Xiaolongetal} Yang, X.,  Joshi, R., Gopal-Krishna, et al. 2019, ApJS, 245, 17

\bibitem[Zwicky \& Herzog(1966)]{ZwickyHerzog} Zwicky, F., Herzog, E., 1966. Catalogue of Galaxies and of Clusters of Galaxies Pasadena, California Institute of Technology

\end{thebibliography}
\end{document}